\documentclass[longauth]{aa}  
\usepackage{graphicx}
\usepackage{txfonts}
\usepackage{hyperref}
\hypersetup{
    colorlinks=true,
    linkcolor=blue,
    filecolor=blue,      
    urlcolor=blue,
    citecolor=blue}
\usepackage{natbib}
\begin{document}
\title{SPICY~30021: an embedded bona fide FUor with a mid-infrared precursor}

%
%
%

 \author{Zhen Guo \inst{1, 2}\fnmsep\thanks{Corresponding author: Zhen Guo, zhen.guo@uv.cl}
           \and
           C. Morris \inst{1}\fnmsep\thanks{Calum Morris, calum.morris@uv.cl}
           \and
           V. Fermiano  \inst{1, 2}
          \and
          C. Lizana-Vidal \inst{1, 2}
          \and         
          A. Aliaga \inst{1, 3}
          \and
          J. Osses\inst{1, 2, 4}
          \and
          F. Navarete \inst{5}
          \and
          P. Lucas \inst{6}
          \and
          R. K. Saito \inst{7}
          \and
          A. Carvalho \inst{8}
          \and
          V. Elbakyan  \inst{9, 10}
          \and
          C. Contreras Peña \inst{11}
          \and
          K. Singh \inst{5}
          \and
          M. Fang \inst{4}
          \and
          P. Weber \inst{12,13}
          \and
          H. Zhao \inst{14, 15}
          \and
          T. Yu. Magakian \inst{16}
          \and
          T. A. Movsessian \inst{16}
          \and
          M. Gomez \inst{15}
          \and
          D. Minniti \inst{15, 17}
          \and
          V. Ivanov \inst{18}
          \and
          M. G. Navarro \inst{19}
          \and
          S. Federle \inst{15}
          \and
          I. Petralia \inst{15}
          \and
          A. Luna \inst{20}
          \and
          J. Corral-Santana \inst{20}
          \and
          E. R. Garro \inst{20}
          \and
          S. D. Points \inst{21}
          \and
          C. Briceño \inst{21}
          \and
          Z. Zhang \inst{22}
          \and
          R. Kurtev \inst{1}
          \and
          J. Borissova \inst{1}
          \and
          J. G. Fernández-Trincado\inst{23}
          \and
          J. Alonso-García\inst{24}
          \and
          C. Cáceres  \inst{15}
          \and
          C. O. Obasi \inst{15}
          \and
          D. Galdeano \inst{25}
           \and
          L. Baravalle \inst{26, 27}
           \and
          T. Giannini  \inst{16}
           \and
          Z. Zhu  \inst{4}
          }
   \institute{Instituto de F{\'i}sica y Astronom{\'i}a, Universidad de Valpara{\'i}so, ave. Gran Breta{\~n}a, 1111, Casilla 5030, Valpara{\'i}so, Chile
         \and
Chinese Academy of Sciences South America Center for Astronomy (CASSACA), National Astronomical Observatories, CAS,
Beijing 100101, China
         \and 
Millennium Nucleus on Transversal Research and Technology to Explore Supermassive Black Holes (TITANS), Valpara{\'i}so, Chile
\and 
Purple Mountain Observatory, Chinese Academy of Sciences, 10
Yuanhua Road, Nanjing 210023, People’s Republic of China
         \and 
Laborat{\'o}rio Nacional de Astrofísica, Rua dos Estados Unidos 154, 37504-364 Itajub{\'a}, MG, Brazil
         \and 
 Centre for Astrophysics Research, University of Hertfordshire, Hatfield AL10 9AB, UK         
         \and 
Departamento de F{\'i}sica, Universidade Federal de Santa Catarina, Trindade 88040-900, Florian{\'o}polis, Brazil
         \and 
Center for Astrophysics — Harvard \& Smithsonian, Cambridge, MA 02138, USA
         \and 
Fakultat fur Physik, Universitat Duisburg-Essen, Lotharstraße 1, D-47057 Duisburg, Germany
        \and
Research Institute of Physics, Southern Federal University, Rostov-on-Don 344090, Russia
          \and 
Department of Physics and Astronomy, Seoul National University, 1 Gwanak-ro, Gwanak-gu, Seoul 08826, Republic of Korea         
\and 
Departamento de F\'isica, Universidad de Santiago de Chile, Av. V\'ictor Jara 3659, Santiago, Chile
\and 
Millennium Nucleus on Young Exoplanets and their Moons (YEMS), Chile
\and 
Institute of Astronomy and Physics, Inner Mongolia University, Hohhot 010021, People's Republic of China
\and
Instituto de Astrof{\'i}sica, Depto. de F{\'i}sica y Astronomía, Facultad de Ciencias Exactas, Universidad Andr{\'e}s Bello, Av. Fernández Concha 700, Las Condes, Santiago, Chile
\and 
Byurakan Observatory NAS Armenia, Byurakan, Aragatsotn Province 0213, Armenia
\and 
Specola Vaticana, Vatican Observatory, V00120 Vatican City State, Italy
\and 
European Southern Observatory, Karl-Schwarzschild-Strasse 2, D-85748 Garching bei München, Germany
\and 
Istituto Nazionale di Astrofisica (INAF), Rome Astronomical Observatory, Via Frascati 33, 00078 Monte Porzio Catone, RM, Italy
\and
European Southern Observatory, Alonso de Córdova 3107, Vitacura, Santiago, Chile
\and
 Cerro Tololo Inter-American Observatory, National Optical Astronomical Observatory, Casilla 603, La Serena, Chile
\and 
School of Astronomy and Space Science, Nanjing University, Nanjing
210023, China
\and
Centro de investigac{\'i}on en Astronom{\'i}a, Facultad de Ingenier{\'i}a, Ciencia y Tecnolog{\'i}a, Universidad Bernardo O'Higgins, Av. Viel 1497, Santiago, 8370993, Chile
\and
Centro de Astronom{\'i}a (CITEVA), Universidad de Antofagasta, Av. Angamos 601, Antofagasta, Chile
\and 
Departamento de Geof{\'i}sica y Astronom{\'i}a, CONICET, Facultad de Ciencias Exactas, F{\'i}sicas y Naturales, Universidad Nacional de San Juan, Av. Ignacio de la Roza 590 (O), J5402DCS, Rivadavia, San Juan, Argentina
\and
 Observatorio Astron{\'o}mico de C{\'o}rdoba, Universidad Nacional de C{\'o}rdoba, Laprida 854, X5000BGR, C{\'o}rdoba, Argentina
 \and
 Instituto de Astronom{\'i}a Te{\'o}rica y Experimental, (IATE-CONICET), Laprida 854, X5000BGR, C{\'o}rdoba, Argentina
 }

 \date{Received xxx; accepted xxx}

  \abstract
   {The unstable mass accretion process on Young Stellar Objects (YSOs) often leads to observable outbursts. These episodic accretion bursts play a key role in the stellar mass assembly during the pre-main-sequence evolution. The highest-amplitude, longest-duration outbursts are commonly identified as FUors. Until recently, only $\sim$50 FUor events have been confirmed. Most of these eruptive events happen in embedded protostars, which are difficult to monitor from optical surveys.}
   {We present photometric and spectroscopic observations of an eruptive young star (SPICY~30021) originally discovered from the mid-infrared (MIR) \textit{NEOWISE} time series. We aim to construct the spectral energy distribution and physically characterise the eruptive nature of SPICY~30021.}
   {We gathered multi-wavelength photometric time series from \textit{NEOWISE}, VVV and the recently established VVV-RENEW surveys. We also conducted photometric and spectroscopic observations during the outburst using instruments on the SOAR, Magellan and VLT telescopes, including data from the ESO public survey VVVX-GalCen. We first identified the location of the outbursting source and disentangled it from nearby companions. We then measured the line-of-sight extinction by comparing spectroscopic features. The stellar parameters were then estimated from radial velocity, proper motions and stellar evolution models.}
   {We found that SPICY~30021 is an optically obscured FUor with a line-of-sight (both interstellar and circumstellar) extinction of $A_V = 12.5\pm0.5$~mag. The spectrum of SPICY~30021 closely resembles those of classical FUors. Using various methods, we measured the distance of SPICY~30021 as $1.7\pm0.4$~kpc. {The accretion luminosity of the outburst is $L_{\rm acc} = 30~\rm L_\odot$, moderate among FUors.} The quiescent near-infrared photometry is consistent with a young, $0.9~\rm M_\odot$ progenitor, although the mass remains model- and distance-dependent.}
   {We confirm SPICY~30021 is a bona fide FUor event. The mid-infrared precursor suggests an outside-in triggering mechanism.}
   
\maketitle
\nolinenumbers
\section{Introduction}

One of the most dramatic stages of pre-main-sequence (PMS) mass assembly is the episodic accretion burst, with large-amplitude outbursts observed at optical and infrared wavelengths \citep{Audard2014, Fischer2023}. During these events, the mass accretion rate temporarily increases by more than 2 dex from the quiescent rate, releasing gravitational energy {that heats} the circumstellar disk, altering its thermal and chemical structure on multiple au scales \citep{Zhu2009, Cieza2016}. The associated heating expands the snowlines of various molecular species, influences dust processing, and triggers time-variable chemistry in the inner few au of protoplanetary disks \citep{Abraham2009, Lee2019NatAs, Jorgensen2020}. Studying PMS outbursts is crucial to reveal how young stars gain their mass, offering direct insight into the physical processes that shape stellar and protoplanetary disk evolution.

The physical mechanism driving these episodic accretion bursts remains uncertain. Theoretical frameworks have invoked a range of instabilities in the accretion disk, including the magnetorotational instability \citep[MRI;][]{Armitage2001, Zhu2009b}, thermal instability \citep[TI;][]{Nayakshin2024b}, or tidal perturbations induced by companions and embedded giant planets on eccentric orbits \citep{Lodato2004, Nayakshin2024}. Outbursts may also arise from the infall or disruption of dense clumps formed by disk fragmentation \citep{Vorobyov2010, Montesinos2026}. In some extreme cases, episodic accretion could
be induced by a stellar flyby to a young stellar system \citep{Cuello2019}.
The FUor outburst is the most drastic event {among} young stellar objects (YSOs), exhibiting optical and near-infrared {(NIR)} brightening by 4–6 mag and lasting for decades \citep{Hartmann1996}. During these outbursts, the accretion mode of the system transitions from magnetospheric to boundary layer. The high accretion rates ($\dot{M}_\mathrm{acc} \sim 10^{-5} \ M_\odot$ yr$^{-1}$) result in a viscously heated inner disk that dominates the system luminosity \citep{Zhu2007, Liu2022, Carvalho2024}. Observationally, the upper disk atmosphere exhibits broad absorption features of CO and H$_2$O and blue-shifted H {\sc i} absorption lines \citep[][]{Hillenbrand2018, Connelley2018}. After the outburst, FUor events have a slow decay in luminosity, tracing viscous draining and cooling of the heated gas reservoir on timescales of decades.

\begin{figure}[!t]
    \centering
    \includegraphics[height=5.08cm]{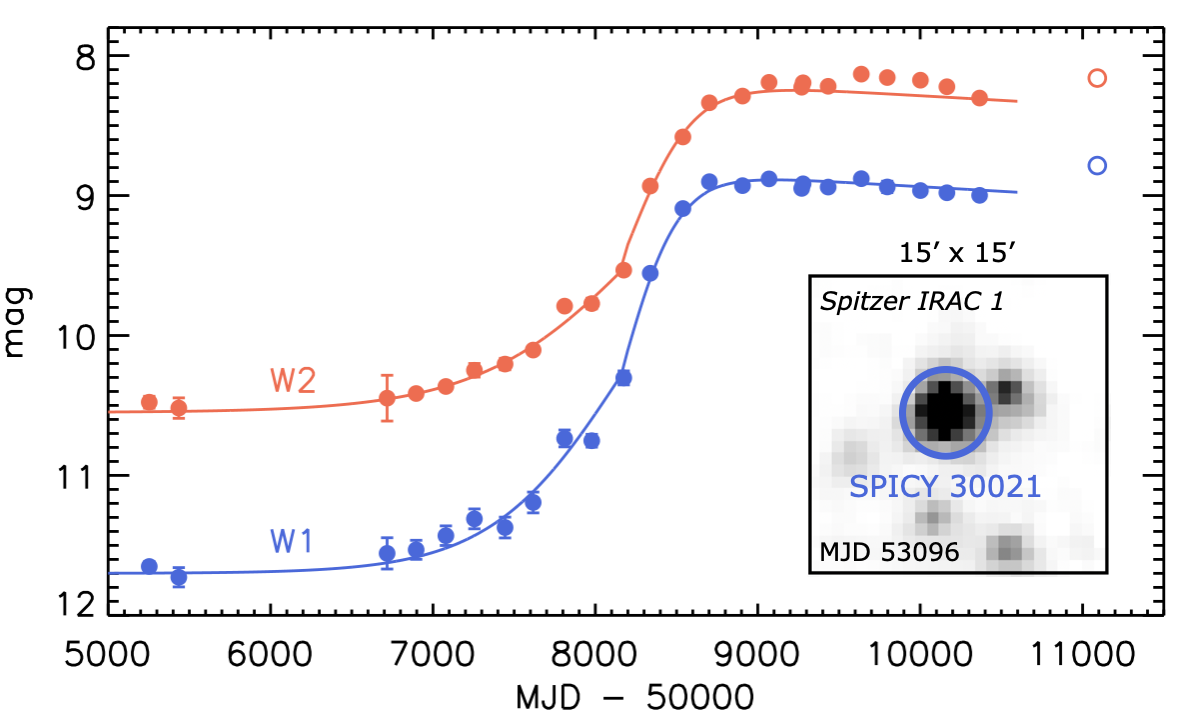}
     \includegraphics[height=5cm]{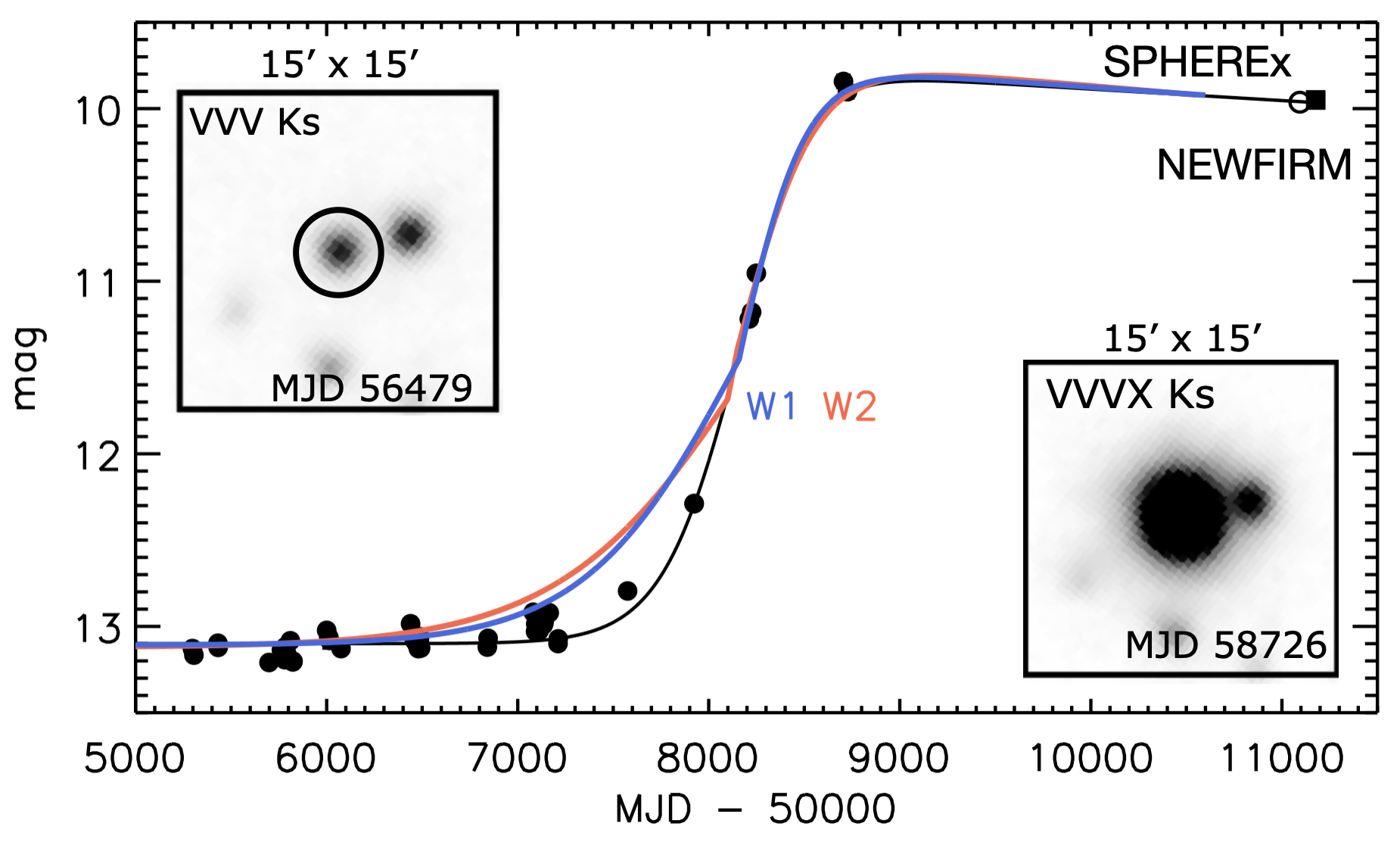}
    \caption{\textit{Upper:} Mid-infrared light curves of SPICY~30021 from \textit{WISE} (filled dots) and \textit{SPHEREx} (open circles near MJD~61090) surveys. We show two analytical functions to fit the rising stages of the outburst. We present a $15' \times 15'$ cutout image from \textit{Spitzer}.  \textit{Lower:} VVV $K_{\rm s}$-band light curve with rising curves. We also present the MIR curve from the {upper} panel, scaled to match the outburst amplitude in $K_{\rm s}$. Two $K_{\rm s}$-band images are shown. The NEWFIRM photometry (filled square) is presented along with the \textit{SPHEREx} data. Error bars are shown when they are greater than the symbol size.}
    \label{fig:MIR_lc}
\end{figure}

Despite decades of observational effort, the population of spectroscopically confirmed FUors remains low (around 50 cases), owing to their rarity and long recurrence timescales ($\sim10^{5}$~yr per star; \citealt{Scholz2013, Hillenbrand2021}). The advent of all-sky time-domain surveys has significantly increased the number of discoveries \citep[see][]{Contreras2025}. Long-term IR photometric monitoring provides a unique diagnostic of accretion bursts, especially towards embedded protostars, where the accretion disk is thought to be unstable. Hundreds of large-amplitude variable young stars were discovered from the decade-long VISTA Variables in the Via Lactea \citep[VVV;][]{Minniti2010} and its eXtension (VVVX) time series \citep{Saito2024}, from 2010 to early 2023 \citep{Lucas2024}, including tens of spectroscopically confirmed FUor objects \citep{Contreras2017b, Guo2020, Guo2021, Guo2024a}. Similarly, the \textit{WISE} mission \citep{Mainzer2014} offers over a decade of coverage in the spectral window of 3–5~$\mu$m, allowing characterisation of the rise and decline of FUors \citep{Li2024, Morris2025, Contreras2025a}. 

Multi-wavelength observation of accretion bursts provides insights into the origin of the instability. {In classical TI models, the outburst is triggered at the innermost regions of the accretion disk (e.g. $r\le 0.1$~au), where the accumulation of material can lead to a thermal runaway associated with hydrogen ionisation \citep{Nayakshin2024b}. The resulting increase in the mass accretion rate sufficiently heats the inner disk, producing an early rise in optical and NIR emission. The enhanced radiation and subsequent heating of cooler, more extended disk regions produce a delayed response at mid-infrared (MIR) wavelengths \citep[e.g.][]{Bell1994, Lodato2004}.}
{In contrast, if the instability is triggered in the outer disk, the initial increase in accretion heating will produce an IR brightening before the optical outburst \citep[e.g.][]{Armitage2001, Cleaver2023}. This is because the cooler disk regions at larger radii contribute significantly to NIR and MIR emission, whilst the optical brightness mainly comes from the innermost disk.} The delayed timescale from optical to IR outbursts also depends on the triggering mechanism. For instance, the dead-zone MRI models predict an IR precursor on timescales from years to decades \citep{Cleaver2023}, whilst the extreme evaporation of a young planet leads to a time delay of months to one year \citep{Nayakshin2024}. Observationally, the IR brightening was detected before the optical {brighting} on a few FUors, indicating outside-in triggering mechanisms \citep{Hillenbrand2018, Guo2024b}. However, the MIR precursor may not be universal, and more multi-wavelength monitored samples are necessary to confirm this scenario. 
 
Here, we present the physical characterisation of a new FUor event (SPICY~30021, RA: 15:51:14.03, DEC: -53:27:49.6, J2000) identified by the VVV and \textit{NEOWISE} time series. It belongs to a list of 12 FUor candidates from the Spitzer/IRAC catalogue of YSOs \citep[SPICY;][]{Kuhn2021}. This source was classified as a flat-spectrum type young star based on the IR spectral energy distribution (SED). In this paper, we confirm its FUor nature and characterise the outburst progenitor using NIR spectra. We also compare the VVV and \textit{NEOWISE} light curves to search for possible time lag between NIR and MIR bands.
In Sect.~\ref{sec:data}, we present the observational data, including two recently established surveys: VVV-RENEW and VVVX-GalCen Spectroscopic Survey. The photometric and spectroscopic results are shown in Sect.~\ref{sec: result}. We compare this event with other known FUors in Sect.~\ref{sec: discussion} and discuss the possible mechanisms causing the MIR precursor. Finally, this paper is summarised in Sect.~\ref{sec: conclusion}. 


\section{Observational data}

\begin{figure*}[!t]
    \centering
    \includegraphics[height=5cm]{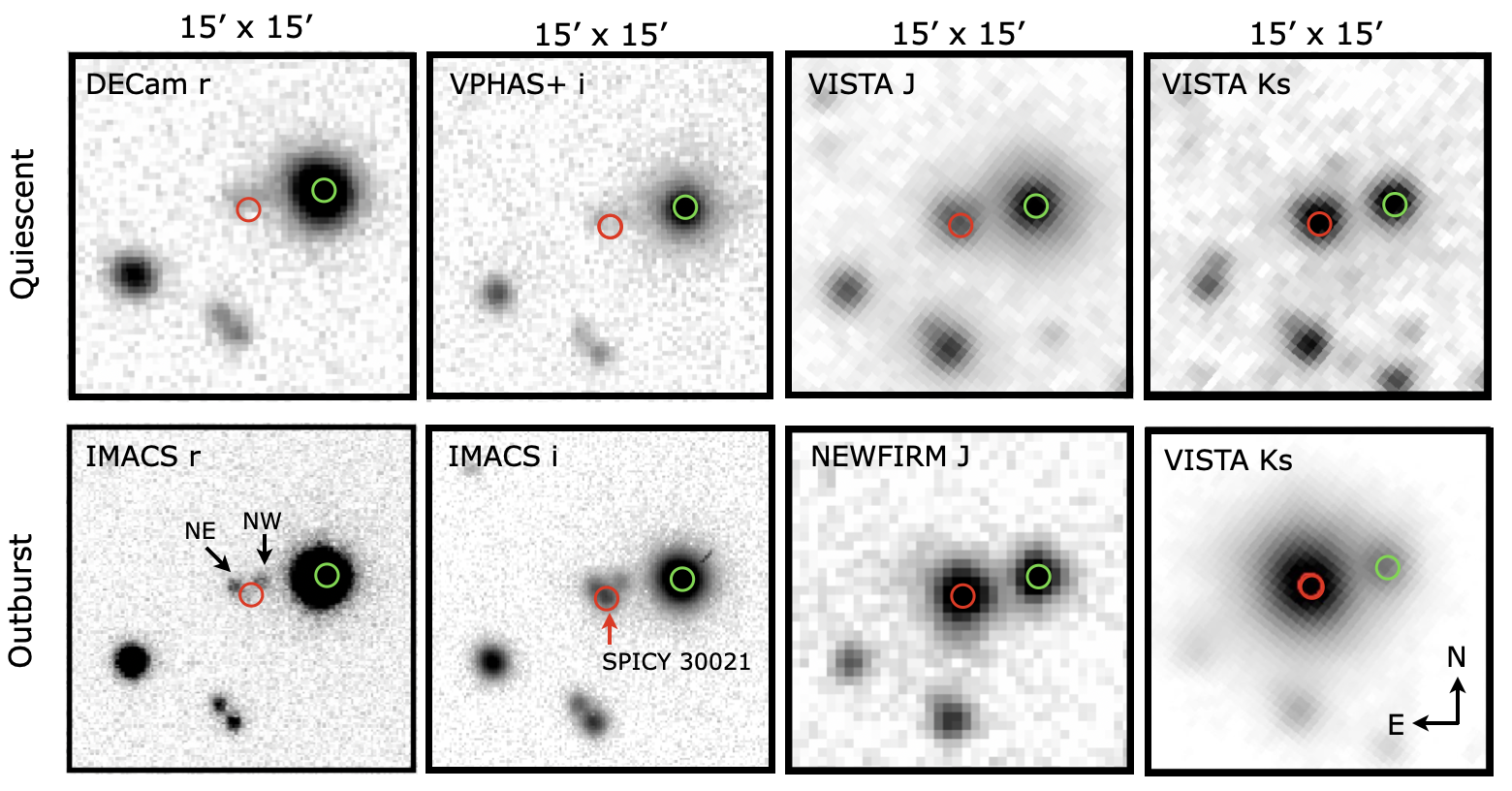}
    \centering
     \includegraphics[height=5.3cm]{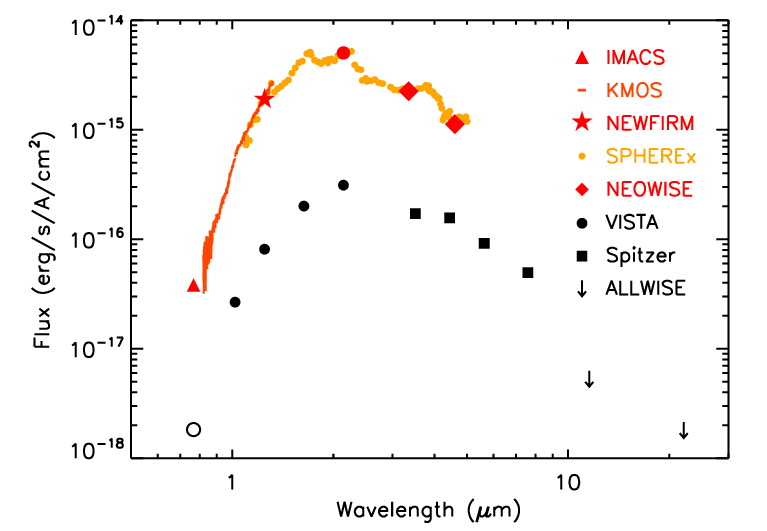}
    \caption{\textit{Left:} Optical to near-IR images of SPICY~30021 (red circle) and the 3\arcsec~companion (green circle) in logarithmic scale. The upper row contains images taken during the quiescent stage, and the outbursting images are shown in the lower row. The North-Eastern and North-Western companions at 1\arcsec are labelled on the $r$-band image. In the $K_{\rm s}$ images, we used different scales between quiescent and outburst phases. \textit{Right:} {Optical ($i$) to mid-IR ($W4$) SED} of SPICY~30021 during the quiescent (black) and eruptive (coloured) stages. The origins of the data points are listed in the legends. A model-predicted $i$-band brightness ($m_i = 22.5$~mag) is shown as an open circle.}
    \label{fig:IMAGE_SED}
\end{figure*}

\label{sec:data}
\subsection{Infrared photometric data}

The eruptive behaviour of SPICY~30021 was identified from the \textit{WISE} time-series photometry, which revealed a brightening of $\Delta W2 > 2$~mag. Owing to the relatively low spatial resolution ($6\farcs1$ -- $6\farcs4$) of the \textit{WISE} data, SPICY~30021 is blended with a nearby companion located at an angular separation of $3\farcs1$. According to the \textit{Spitzer} GLIMPSE catalogue \citep{Churchwell2009}, SPICY~30021 has magnitudes of $I1 = 11.50\pm0.04$~mag {(IRAC 3.6~$\mu$m)} and $I2 = 10.61\pm0.05$~mag {(IRAC 4.5~$\mu$m)}, whereas the companion has $I1 = 13.00\pm0.05$~mag and $I2 = 12.79\pm0.12$~mag. Between \textit{Spitzer} and \textit{WISE} filters, disk-bearing YSOs usually have mean $W1 - I1 = 0.2$ and $W2 - I2 = -0.1$ mag (Singh et al., in prep.), which is on the same order as the photometric uncertainties of the \textit{NEOWISE} time series. Therefore, we assumed $I1 \approx W1$ and $I2 \approx W2$, and that the companion is non-variable; we then corrected the \textit{WISE} photometry by subtracting the companion's flux. The resulting light curve, shown in Fig.~\ref{fig:MIR_lc}, combines observations from the \textit{ALLWISE}, \textit{WISE} 3-Band Cryo, and \textit{NEOWISE} surveys, providing a cadence of approximately six months. Individual measurements obtained within the same observing epoch were averaged to improve the signal-to-noise ratio. For reference, a \textit{Spitzer} image obtained during the quiescent phase is shown in Fig.~\ref{fig:MIR_lc}, illustrating the positions of SPICY~30021 and its $3\farcs1$~companion.

We obtained multi-band NIR time-series photometry from the VVV VIRAC2 catalogue \citep{Smith2025}. During the outburst maximum (around MJD~58700), the $K_{\rm s}$ brightness approached the saturation limit of the VISTA detectors ($K_s \sim 10$~mag). We therefore excluded data points with $\rm ast\_res\_chisq > 10$, as these are likely affected by saturation. For epochs with $K_s < 11$~mag, we corrected the non-linear photometry using a procedure that recovers the saturated flux by fitting and scaling a point-spread function (PSF) model from nearby non-saturated sources in the same field-of-view (Navarete et al., in prep.). The resulting correction factors are typically on the order of $-0.05$~mag.  We also retrieved PSF photometry in the $Y$, $J$, $H$, and $K_{\rm s}$ bands from the VIRAC2 catalogue. Visual inspection of the images revealed two bluer sources within 1\arcsec~of SPICY~30021 in the $Y$-band (and also in the $Z$-band, where SPICY~30021 itself is not detected). Consequently, the $Y$-band photometry may be contaminated by these nearby sources.

In May 2026, we obtained $JX$, $HX$, and $KX_s$-band imaging of SPICY~30021 as part of the VVV-RENEW project, a NIR survey conducted with the NEWFIRM imager \citep{Autry2003} on the 4-m Blanco Telescope (PI: Guo). The NEWFIRM imager has a field of view of $28\arcmin \times 28\arcmin$ with a pixel scale of $0\farcs43$. The VVV-RENEW survey covers approximately 70~deg$^2$ of the Southern Galactic plane and is designed to achieve a dynamic range of $JX$, $HX$, and $KX_s \simeq 8$--18~mag. For each field, we co-added two 2-second exposures at each pointing and repeated this sequence over a 12-point dither pattern to fill detector gaps and mitigate bad pixels. The data were reduced using a customised version of the NEWFIRM pipeline developed by Dr Sean Points. A full description of the survey and its data product will be presented in a forthcoming publication (Fermiano et al., in prep.). The measured brightness of SPICY~30021 on May 17 2026 was $JX = 12.97 \pm 0.02$~mag and $KX_s = 9.95 \pm 0.01$~mag.

We retrieved \textit{SPHEREx} spectra \citep{Bock2026} from the NASA/IPAC IR Science Archive (IRSA)\footnote{\url{https://irsa.ipac.caltech.edu/}}. \textit{SPHEREx} is a NASA space telescope launched in 2025 that is conducting an all-sky near- and mid-IR spectrophotometric survey with a spatial resolution of $6\farcs15$. The survey provides low-resolution spectra in 102 spectral channels covering the wavelength range 0.75--5.0~$\mu$m. Two \textit{SPHEREx} spectra are currently available for SPICY~30021, and no significant variability ($>0.1$~mag) is detected between the two epochs. Owing to the relatively low spatial resolution, SPICY~30021 is blended with nearby sources, particularly at shorter wavelengths. In Appendix~\ref{sec:spherex}, we present the SED of the nearby companion and describe the spectrophotometric extraction procedure. We also integrated $K_{\rm s}$, $W1$, and $W2$ photometry from the \textit{SPHEREx} data over the corresponding filter transmission curves, and included these measurements in the multi-epoch light curve shown in Fig.~\ref{fig:MIR_lc}.

\subsection{Optical photometric data}

\begin{figure*}
    \centering
    \includegraphics[height=6cm]{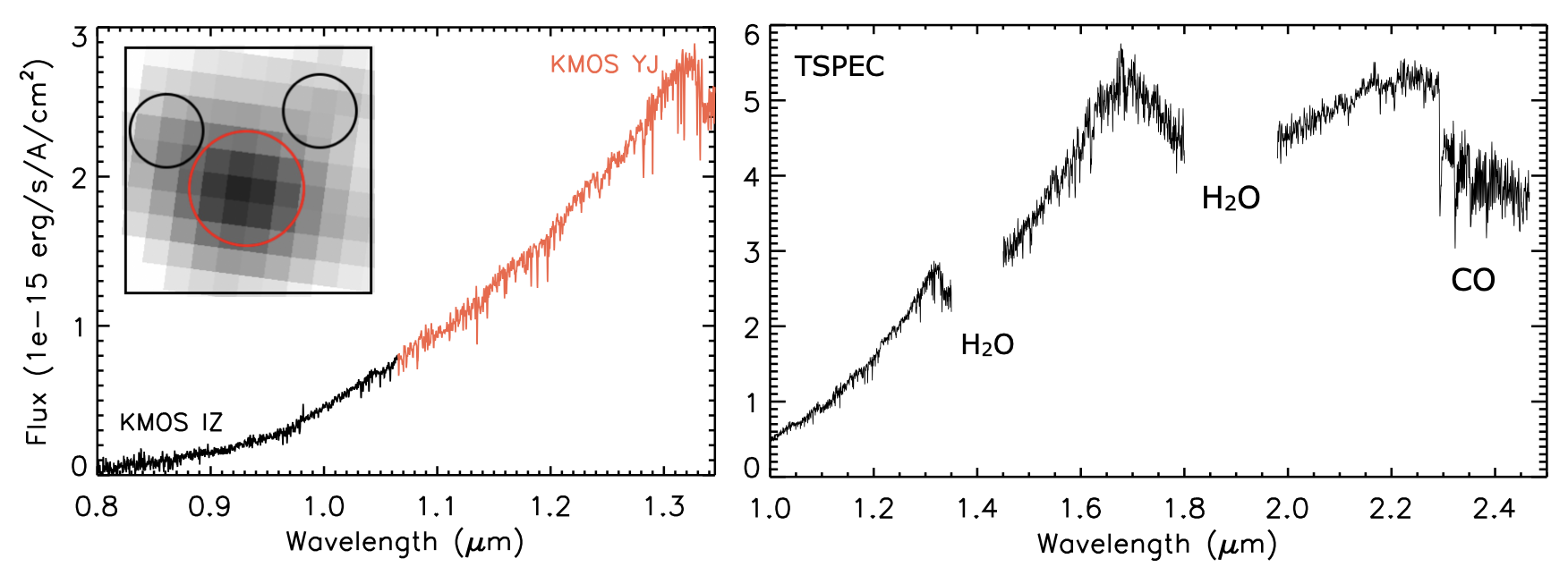}
    \caption{Near-IR spectra of SPICY~30021 during the outburst. \textit{Left:} KMOS $IZ$ (black) and $YJ$-band (red) spectra, observed on May 17 2026 and May 18 2026, respectively. The 2D $IZ$-band spectral image is presented as a subplot, with {SPICY~30021} marked as the red circle. \textit{Right:} The TSpec spectrum observed on May 23 2026, with spectral features marked.}
    \label{fig:spec}
\end{figure*}
We retrieved three pre-outburst $i$-band images from the VPHAS+ survey \citep{Drew2014} through the ESO Science Archive. The images were obtained in 2012, well before the onset of the outburst in September 2016. The co-added image reaches a $5\sigma$ limiting magnitude of $i = 22.0$~mag. No source is detected at the position of SPICY~30021. We also retrieved an $r$-band image obtained with DECam in 2016, a few months before the outburst. This image is part of the Dark Energy Camera Plane Survey (DECaPS) DR2 \citep{Saydjari2023}, which reaches a designed depth of $r \simeq 22.6$~mag. As in the $i$-band, SPICY~30021 was not detected.

On May 8 2026, we obtained optical images of SPICY~30021 with the IMACS instrument on the 6.5-m Magellan Baade Telescope at Las Campanas Observatory (PI: Morris). The IMACS f/4 camera has a pixel scale of $0\farcs11$, enabling us to resolve nearby companions to SPICY~30021. The observations were carried out with exposure times of 60~s in $i$ and 120~s in $r$. From the $i$-band image, we measured an apparent magnitude of $i = 19.26 \pm 0.05$~mag. Two nearby companions are resolved approximately 1\arcsec~north of SPICY~30021. No source is detected at the target position in the $r$-band image at the $3\sigma$ level, consistent with the object being heavily affected by extinction. The North-Eastern companion has magnitudes of $r = 21.7 \pm 0.2$~mag and $i = 19.7 \pm 0.1$~mag, while the North-Western companion has $r = 21.8 \pm 0.2$~mag and $i = 20.5 \pm 0.1$~mag. These two close-in companions are also seen in the VISTA $J$-band image before the outburst and could potentially contaminate the MIR images, although they are much fainter than SPICY~30021 at longer wavelengths. Individual reduced images are shown in Fig.~\ref{fig:IMAGE_SED}.

\begin{figure*}
\centering
\includegraphics[width=\linewidth]{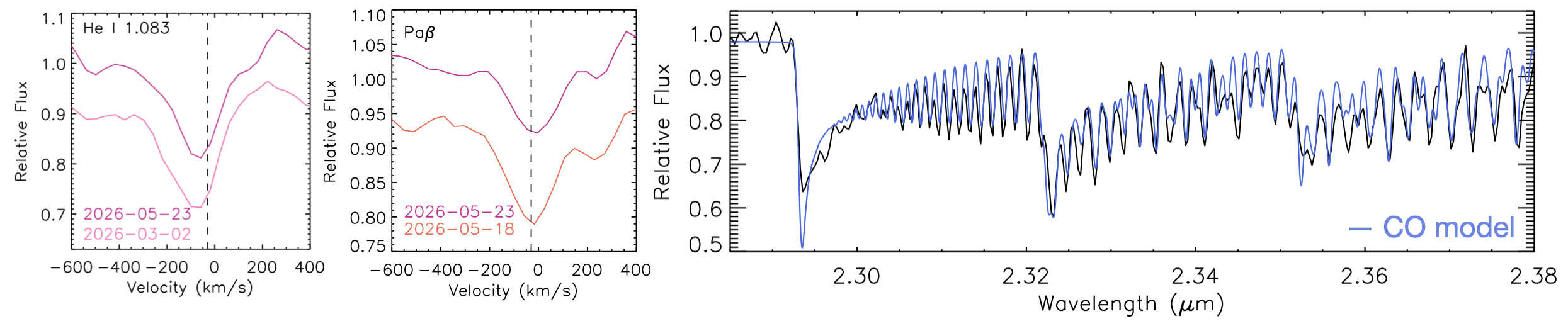}
\caption{\textit{Left} and \textit{Middle}: Absorption line profiles of He {\sc i} and Pa $\beta$ with data from TSpec (March 2 2026 and May 23 2026) and KMOS (May 18 2026). The vertical dashed line marks the RV of CO bandheads. \textit{Right}: CO bandhead absorption spectrum of SPICY~30021 (black, May 23 2026). We normalised the continuum flux between 2.28 and 2.45 $\mu$m. Our CO model is presented in blue.}
\label{fig:CO}
\end{figure*}


\subsection{Spectroscopic data}
We observed four NIR spectra of SPICY~30021 using TSpec and KMOS spectrographs. Two epochs of NIR spectra were obtained by the TSpec spectrograph \citep{Schlawin2014} mounted on the SOAR telescope, on the nights of March 2 2026 (PI: Fermiano) and May 23 2026 (PI: Guo). The observations were conducted in the ABBA nodding sequence with 300~s per single exposure. In each run, an A0V telluric standard star was observed immediately after the science target for telluric correction. The data were reduced by the IDL-based reduction pipeline including 1D spectrum extraction, telluric correction and merging of spectral orders. We then carried out a customised procedure to {correct} the solar barycentric motion.
We note that no absolute flux calibration was done to compensate for the slit loss. Therefore, we scaled the reduced spectrum to match the broadband photometry from NEWFIRM observed on May 17 2026, as an external flux calibration. 

Two NIR spectra were observed by the KMOS spectrograph mounted on the UT1 of the Very Large Telescope (VLT) at Paranal Observatory, as part of the ESO Public Spectroscopic Survey VVVX-GalCen \citep[][and Gomez et al., in prep.]{Arnaboldi2026}. KMOS is an integral field spectrograph that enables 24 integral field units (IFUs) of $14\times14$ spatial pixels with a {sampling} of $0\farcs2 \times0\farcs2$ each, covering a field of $2\farcs8 \times2\farcs8$ per IFU \citep{Sharples2004}. {The designed spectral resolution ranges between $R=3400$ in $IZ$ and 3600 in $YJ$}. We observed SPICY~30021 in the $IZ$-band on May 17 2026, and in the $ YJ$-band on May 18 2026. After co-adding and dithering, the total integration time is 25 min in each band.  We reduced the spectra using the KMOS pipeline on the ESO Data Processing System \citep[][]{Freudling2024} and utilised Molecfit for telluric correction \citep{Smette2015}. Using the QFitsView software \citep{Ott2012}, we extracted the spectrum of SPICY~30021 from the 2D spectral image using a 3-pixel aperture ($0\farcs6$) and subtracted the background emission over a 6-pixel aperture. The extracted spectra and spectral images are presented in Fig.~\ref{fig:spec}.

\section{Results}
\label{sec: result}
\begin{figure*}
    \centering
    \includegraphics[width=\linewidth]{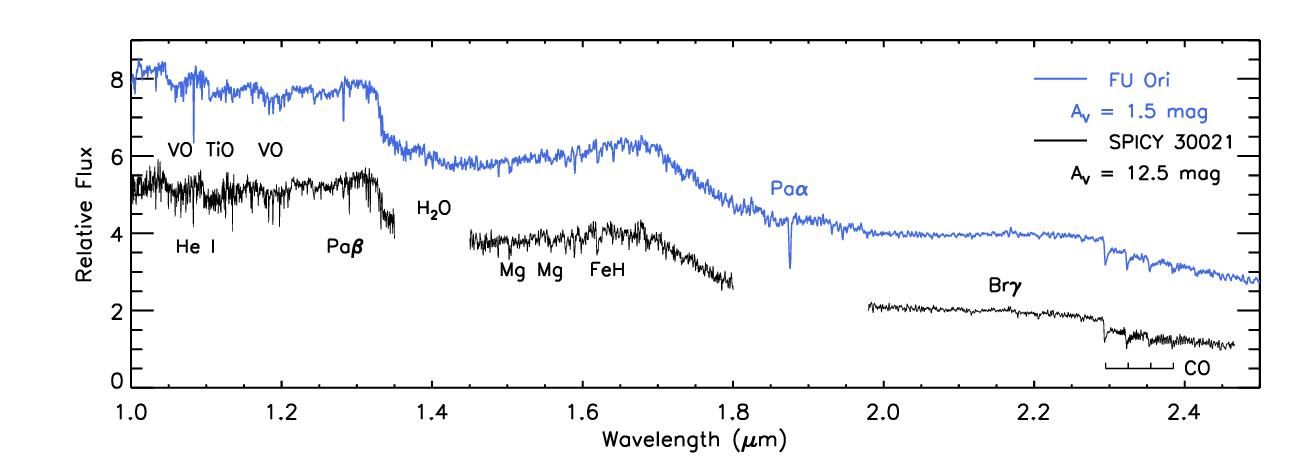}
    \caption{De-reddened NIR spectra of SPICY~30021 and FU Ori \citep[][]{Connelley2018}, using the reddening law from \citet{Cardelli1989}.}
    \label{fig:spec_compare}
\end{figure*}

\subsection{Photometric variability}

The earliest measurement of SPICY~30021's $K_{\rm s}$-band brightness is $13.05\pm0.06$~mag, observed in June~1999 by the 2MASS survey \citep{Cutri2003}. In the VVV survey era, SPICY~30021 exhibited low-amplitude variations (0.1~mag in $K_{\rm s}$) until late 2015. In September 2016, a brightening trend was observed in the VVV survey data, which reached a photometric maximum in {August 2019}. The saturation-corrected VVV/VVVX light curve yields a peak-to-peak amplitude of $\Delta K_{\rm s} = 3.3$~mag, although only two data points were obtained near the photometric maximum. As mentioned in Sect.~\ref{sec:data}, we assumed the 3\arcsec~companion is a non-variable source and removed its contribution from the \textit{NEOWISE} time series. We measured MIR amplitudes of SPICY~30021 as $\Delta W1 = 2.8$~mag and $\Delta W2 = 2.4$~mag, smaller than $\Delta K_{\rm s}$. We found that before entering the rapid brightening stage {along with the NIR (MJD 58200)}, the MIR brightness increased gradually for more than 1500 days. Until mid-2026, SPICY~30021 was still situated on the brightness plateau for more than six years (see Fig.~\ref{fig:MIR_lc}), {consistent with the expected long duration of a FUor-type event.}

To further illustrate the different rising slopes between NIR and MIR bands, we fit analytical functions to the light curves, following the methods introduced by \citet{Lucas2024}. The fits indicate that the MIR emission, particularly in the $W2$-band, began to brighten approximately 900~days before the onset of the $K_{\rm s}$-band rise (MJD~57600). The characteristic rising timescales $\tau$ in  $K_{\rm s}$, $W1$, and $W2$ bands are 170, 400 and 500~d, respectively. This behaviour is consistent with the outside-in outburst scenario reported for several other FUors \citep{Hillenbrand2018, Szegedi-Elek2020, Contreras2026}. However, the earlier brightening of the MIR bands may reflect the temperature evolution of the inner disk, rather than the location of the outburst \citep{Das2026}. Once the outburst had reached approximately half of its total amplitude, however, the three IR bands exhibited nearly identical rising behaviour, consistent with the propagation of heating through the accretion disk following the onset of the eruptive event.

\subsection{Spectroscopic features}

The IR spectrum of an FUor is characterised by a luminous accretion disk that outshines the stellar photosphere, resembling a late-type object. Recent spectroscopic studies have shown that bona fide FUors exhibit similar spectral characteristics \citep[][]{Connelley2018}. Figures~\ref{fig:spec} to \ref{fig:spec_compare} present the 0.80--2.45~$\mu$m spectrum of SPICY~30021, obtained by KMOS and TSpec. The spectra display characteristic features of an FUor disk, including strong H$_2$O absorption bands, CO overtone bandhead in absorption, Pa~$\beta$ absorption line, numerous metal lines (Mg~{\sc i}, Si~{\sc i}, Ca~{\sc i}, etc.), and blueshifted He~{\sc i} absorption line.  Zoomed-in views of the NIR spectra are presented in Appendix~\ref{sec:line}.

Among various spectral features resembling classical FUors, we present the line profiles of  He~{\sc i} $\lambda$1.083~$\mu$m and Pa~$\beta$. Both lines are broad with FWHM near 250~km~s$^{-1}$, greater than the spectral resolution ($R = 3500$). The equivalent widths (EWs) of He~{\sc i} and Pa~$\beta$ lines are 2.35~and 0.87\AA, respectively. When compared with CO bandheads, the He~{\sc i} line is asymmetric and blue-shifted by 40~km~s$^{-1}$, resembling a stellar wind or outflow. On the contrary, Pa~$\beta$ is found symmetrically located around 0~km~s$^{-1}$, {suggesting a disk origin rather than a stellar wind. }

FUors and late-type stars share very similar overall spectral features in the NIR, including molecular and atomic absorption lines. However, they are distinguishable by the EW of individual spectral features \citep{Connelley2018, Portnoi2026}, including Pa~$\beta$, Na~{\sc i}, and Ca~{\sc i} lines. The Pa~$\beta$ EW (0.87\AA) lies in the overlapping region occupied by FUors and young brown dwarfs \citep{Connelley2018}. {In addition, the EWs of Na~{\sc i} $\lambda$2.21~$\mu$m and Ca~{\sc i} $\lambda$2.26~$\mu$m are 1.29 and 3.40~\AA}, respectively. These values are among the largest EWs reported for FUors \citep[see][]{Portnoi2026}, yet they remain smaller than those measured in late-type stars. We fitted a molecular slab model to the CO overtone bandhead absorption at wavelengths beyond 2.29~$\mu$m, using the spectrum obtained with TSpec in May 2026. The models were generated using molecular transition probabilities and partition functions from the ExoMol database \citep{Tennyson2016, Yurchenko2018}, with the effective temperature ($T_{\rm eff}$), CO column density ($N_{\rm CO}$), and radial velocity ($\rm RV$) treated as free parameters. A more detailed description of the modelling methodology is presented in \citet{Guo2026}. As noted by \citet{Contreras2017b}, the current CO models do not fully reproduce the broadening of the first bandhead at 2.29~$\mu$m. Nevertheless, the RV can still be constrained with a precision ($\sim$5~km~s$^{-1}$) exceeding the nominal spectral resolution. The best-fitting model yields $V_{\rm LSR} = -26.5$~km~s$^{-1}$, $T_{\rm eff} = 2000$~K, and $N_{\rm CO} = 7\times10^{20} \,\rm cm^{-2}$ (Fig.~\ref{fig:CO}). The $T_{\rm eff}$ and $N_{\rm CO}$ are similar to other IR-discovered FUors \citep[see][]{Contreras2017b, Guo2021}.
Previous work reported a case of {a blueshifted CO wind }arising in one individual eruptive YSO \citep{Carvalho2024b}, which would affect the measurement of the systematic RV if the CO wind is universal among FUors.
 
To estimate the line-of-sight extinction ($A_V$) toward SPICY~30021, we followed the method of \citet{Guo2024a} by matching its $J$- and $H$-band spectral continua to those of FU~Orionis, the prototype of the FUor class. We excluded the $K$-band spectrum from this comparison because several FUors exhibit excess emission at these wavelengths, attributable to the dust disk \citep[see Fig.~1 of][]{Guo2024a}. The FU~Ori spectrum was obtained with IRTF and dereddened by $A_V = 1.5$~mag according to the measurements in \citet{Connelley2018}. We adopted the reddening function from \citet{Cardelli1989} and assumed $R_V = 3.1$. We then applied the same reddening function with $A_V$ ranging from 5 to 15~mag to SPICY~30021. By comparing the $J$ and $H$-band continuum slope between SPICY~30021 and FU~Ori (1.0 to 1.8 $\mu$m) through visual inspection, we found the best-fitting extinction $A_V$ is 12.5 mag. The dereddened spectra show a remarkable similarity in both continuum and spectral features (Fig.~\ref{fig:spec_compare}).  However, we comment here that FUors may exhibit intrinsic diversity in disk temperature, inclination and scattered light. {We therefore adopt a systematic uncertainty of 0.5 mag for the best-fit value (see Appendix for de-reddened spectra with different $A_V$). }

In summary, we classify SPICY~30021 as a bona fide FUor as it satisfies the following photometric and spectroscopic criteria. It had a {3.4 mag outburst in $K_s$} and lasted more than 2000 days. We identified spectral features that resemble a cool but luminous accretion disk, including broad water vapour absorption, CO overtone absorption, and other absorption lines. We identified a blueshifted He~{\sc i} absorption line {indicative of a stellar wind. }

\subsection{Stellar properties of SPICY~30021}

We estimated the distance of SPICY~30021 by applying the following methods. First, we calculated the distance using the proper motions (PMs) of SPICY~30021 derived from VIRAC2, resulting in a distance of $d_{\rm PM} = 1.6\pm0.5$~kpc. Second, using the RV measured from the CO bandheads ($V_{\rm LSR} = -27$~km~s$^{-1}$), we calculated a near kinematic distance $d_k = 1.7\pm0.4$~kpc from the Galactic rotational models designed in \citet{Wenger2018}. Details of the distance measurements are presented in Appendix~\ref{sec:distance}. The $d_{\rm PM}$ and $d_{\rm k}$ measured here are consistent with each other and also in line with the location of the Sagittarius–Carina arm towards $gl = 327.8^\circ$ (1.6-1.8 kpc, see Appendix). In the following studies, we will use $d = 1.7\pm 0.4$~kpc as the distance to SPICY~30021.

We then calculated the accretion disk luminosity during the outburst, applying the bolometric correction for FUors \citep{Carvalho2024}. Adopting $d=1.7$~kpc and $A_V = 12.5$~mag, we measured $M_J = -1.15$~mag from the NEWFIRM image, with a 0.6~mag {uncertainty} on the distance modulus. The uncertainty introduced by the extinction measurement {(e.g. $\delta A_J = 0.1$~mag) } is smaller than that from the distance. Applying the $BC_J$ derived in \citet{Carvalho2024}, we calculated the bolometric luminosity as 
\begin{equation}
    L_{\rm bol} = 10 ^ {-0.4\times(M_J + BC_J - M_{\rm bol, \odot})}\,\rm L_{\odot},
\end{equation}
where $BC_J = 2.20$ mag and $M_{\rm bol, \odot} = 4.74$ mag. {The calculated disk luminosity of SPICY~30021 is $L_{\rm bol} \sim30\, \rm L_\odot$. When accounting for distance error (0.4~kpc), the estimated luminosity ranges from 18 to 46 $L_\odot$. } We note here that this luminosity is derived from an accretion disk model, not including flux from the stellar photosphere. 

{Applying $A_V = 12.5$~mag and $d = 1.7$~kpc, we measured the quiescent $J$- and $H$-band brightness as $M_J = 2.20$~mag and $M_H = 1.58$~mag. Both $M_J$ and the ($J-H$) colour index are consistent with a 0.9~M$_\odot$ young star at an age of less than 1~Myr, according to the BHAC15 models \citep{Baraffe2015}. If SPICY~30021 has $d = 2.1$~kpc, the estimated stellar mass is 1.2-1.3~M$_\odot$, a mass range in which FUor outbursts are rarely detected. Following the same logic, we exclude the possibility that SPICY~30021 is located at the far kinematic distance (12~kpc). Alternatively, if SPICY~30021 is located at 1.3~kpc, closer than we estimated, it would be a low-mass YSO with $M_\star < 0.5$~M$_\odot$. In addition, we note that the mass and age estimation is heavily model-dependent, and any unknown stellar variability will bias the estimated values.}


{The estimated $L_{\rm bol}$ allows an order-of-magnitude estimate of the burst properties. We assume that $L_{\rm acc}$ equals the disk $L_{\rm bol}$ and is released at the stellar surface, $L_{\rm acc}=GM_\star\dot M/R_\star$. The observed value of $L_{\rm acc}\approx30 \rm \,L_\odot$, together with $M_\star\approx0.9 \rm \,M_\odot$ and a stellar radius of $R_\star\simeq 2.9 \rm \,R_\odot$, implies an accretion rate of $\dot M\simeq3.1\times10^{-6} \rm \, M_\odot\, {\rm yr^{-1}}$. Considering the distance uncertainty, the measured mass accretion rate is $\dot M\simeq4.3\times10^{-6} \rm \, M_\odot\, {\rm yr^{-1}}$ when assuming a large distance ($d = 2.1$~kpc) and $\dot M\simeq2.9\times10^{-6} \rm \, M_\odot\, {\rm yr^{-1}}$ when applying $d = 1.3$~kpc.
} Given the uncertainties in distance, extinction, disk inclination, and the appropriate inner radius, this value should be regarded as an order-of-magnitude estimate. If the enhanced accretion persists for approximately a few decades, the total mass accreted during the burst is of order $10^{-4} \rm \,M_\odot$, consistent with the FUor accretion regime. From the beginning of the outburst until mid-2026, about 7~$M_\oplus$ {have been accreted onto the young star. }

\section{Discussion}
\label{sec: discussion}

\begin{figure}
\centering
\includegraphics[width=\linewidth]{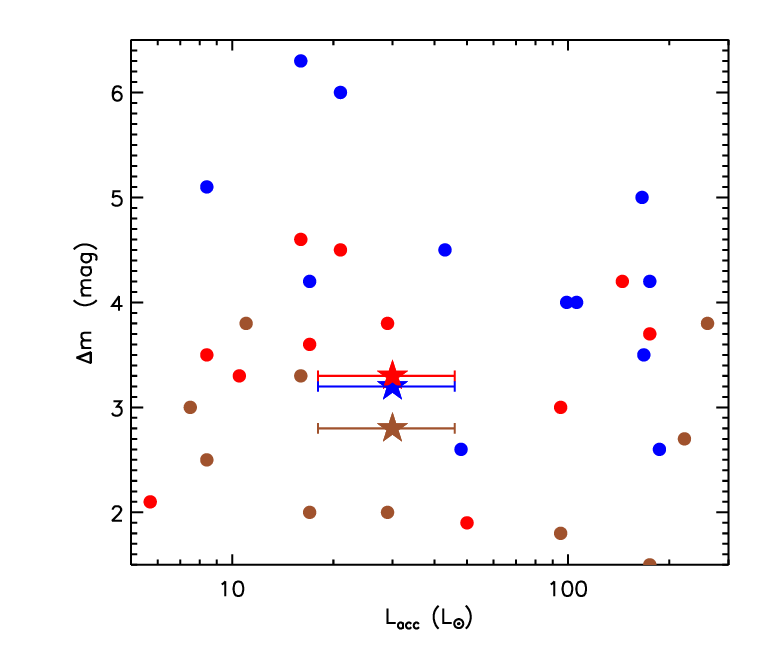}
\caption{$W2$ (brown), $K_{\rm s}$ (red) and optical (blue) amplitudes versus $L_{\rm acc}$ of bona fide FUors that have robust distance measurements. SPICY~30021 is shown as stars with error bars due to distance uncertainties, while other FUors are presented as coloured dots. The optical amplitude of SPICY~30021 is in the $i$-band. Most optical amplitudes of other FUors are in the $R$- or $G$-band.}
\label{fig:FUors}
\end{figure}

The FUor outburst is commonly defined as a sudden and dramatic increase in the optical to IR luminosity of a YSO, typically by 4--6~mag over a short timescale, followed by a slow decay lasting several decades \citep{Hartmann1996, Fischer2023}. However, the growing number of bona fide FUors and FUor-like sources identified in large-scale variability surveys---particularly in the IR---has broadened the empirical classification criteria for these events. Calculated from the difference between the quiescent and eruptive SED, SPICY~30021 exhibits amplitudes\footnote{Adopting the $i$-band brightness from BHAC models. The observational lower limit is $\Delta i > 2.7$ mag considering the non-detection from the quiescent phase.} of $\Delta i = 3.3$, $\Delta J = 3.4$, $\Delta K_s = 3.3$, $\Delta W1 = 2.8$ and $\Delta W2 = 2.4$~mag. In Fig.~\ref{fig:FUors}, we compare these amplitudes and corresponding $L_{\rm acc}$ with a sample of confirmed FUors from the OYCAT catalogue \citep{Contreras2025}. The optical-to-NIR amplitude of SPICY~30021 lies toward the lower end of the classical FUor distribution, comparable to objects such as Gaia20bdk and V7995~Sgr \citep{Siwak2025, Hillenbrand2026}. On the other hand, the $L_{\rm acc}$ of the outburst in SPICY~30021 is moderate on the log scale among bona fide FUors. 

The case of SPICY~30021, as a relatively low-amplitude but moderate-luminosity FUor, highlights the growing complexity in searching for and identifying subcategories of episodic accretion events. The wide diversity of amplitudes and eruptive timescales revealed by time-domain surveys suggests that FUors may represent one extreme of a continuous distribution of burst intensities, rather than a discretely defined class. {Spectroscopic follow-up of variable sources in star-forming regions is therefore essential;} it has already revealed a significant number of extremely unusual sources associated with strong outflows or outbursts \citep[e.g.,][]{Movsessian2021, Movsessian2025, Guo2025, Magakian2026}. In addition, the stellar photospheric brightness affects the eruptive amplitude, as more massive young stars exhibit lower eruptive amplitudes at the same $L_{\rm acc}$. For the same reason, accretion bursts are difficult to detect on intermediate or massive YSOs \citep{Elbakyan2025}.

The observation of SPICY~30021 offers valuable insight into the physical processes driving FUor outbursts. The gradual MIR brightening preceding the more rapid NIR rise favours an outside-in heating sequence, suggesting that the instability originated in the cooler, more extended region of the disk rather than in the magnetospheric boundary. This interpretation is consistent with the gradual rise profile observed in SPICY~30021 and other FUors with similar pre-outburst MIR activity \citep[e.g., Gaia17bpi;][]{Hillenbrand2018}. TI models predict that once a local enhancement of surface density or temperature exceeds the critical opacity threshold at the innermost disk, ionisation feedback triggers a runaway increase in viscosity, and the ionisation front then sweeps outwards \citep{Bell1994, Zhu2007}. The early MIR response might trace the outward propagation of the heating front, while the delayed $K_{\rm s}$-band brightening corresponds to the inward migration of material toward the stellar magnetosphere. However, the TI region is confined to $r \lesssim 0.1$\,au \citep{Bell1994}, which may be too compact to produce a s{ignificant time delay ($> 1$~yr)} among $K_{\rm s}$-, $W1$- and $W2$-band light curves. By introducing an evaporating planet, \citet{Nayakshin2024} were able to produce a MIR precursor on a timescale of months to about a year. In this model, the planet triggers a TI in the disk and subsequently undergoes extreme evaporation, supplying additional mass to the inner disk and sustaining the accretion burst. 

On the other hand, activation of the MRI at the inner edge of the dead zone produces an inside-out burst, with little or no lag between bands \citep{Cleaver2023}.{ However, the MRI could be activated further out by significant mass loading. In this case, the dead zone becomes thermally ionised and leads to an increase in viscosity. A boundary is therefore drawn between the hot/MRI-active region outwards and the quiescent disk inwards \citep{Cleaver2023}. The accretion outburst occurs when this boundary reaches the star, and while it is still traversing the disk, a linear rise is detectable at IR wavelengths alone.} This front would advance much faster than the gas itself drifts inward \citep{Cleaver2023}, and the MIR is expected to lead the optical and NIR by years to decades \citep{Armitage2001, Zhu2009b, Bae2014, Cleaver2023, Nayakshin2024}, compatible with the $\sim$900\,d $W2$-band precursor of SPICY~30021.

As demonstrated by \citet{Lucas2024}, variable sources are ten times more numerous when the detection threshold is lowered by one magnitude in $\Delta K_s$. However, the essential spectroscopic confirmation remains severely constrained by telescope time. In the near future, the wide wavelength coverage {(UV to NIR)} and deep imaging capability of the Rubin Observatory's LSST will enable the discovery of hundreds of episodic accretion events each year, including some moderately embedded sources that become optically detectable during outbursts. The recent development of machine-learning–based community brokers will further facilitate the identification of such transients, though a morphological and multi-wavelength classification framework is likely more robust than simple amplitude thresholds \citep{Alerce2021}. Moreover, regular-cadence IR monitoring with \textit{SPHEREx}, the VVV-RENEW survey, and the Roman Galactic Plane Survey \citep{Roman2025} will provide a powerful diagnostic for detecting and confirming the earliest eruptive stages, in some cases even before substantial optical brightening.

\section{Conclusion}
\label{sec: conclusion}

In this work, we present a comprehensive analysis of the new FUor outburst SPICY~30021, combining multi-epoch optical and IR data. The source exhibits classical signatures of an FUor eruption, including a drastic increase in brightness ($\Delta K_{\rm s} = 3.3$~mag), a long-duration outburst plateau and a late-type NIR spectrum. We summarise our discovery as follows.

\begin{itemize}
    \item We discovered an IR eruptive event in the YSO SPICY~30021 that started in 2016 (around MJD 57600). The outburst reached its maximum brightness approximately three years after the initial IR brightening. We identified a gradual increase in the MIR light curve nearly 900 days before the NIR eruption, consistent with an outside-in triggering mechanism, potentially involving MRI activation. {The $W2$-band brightness rises slightly earlier than $W1$, although no significant difference is observed despite the amplitude difference. }

    \item By inspecting the images, we identified multiple objects located within 3\arcsec~from our target, including two faint sources blended with SPICY~30021 in the optical and NIR images. During the outburst, the brightness of SPICY~30021 dominates in the NIR. However, it was not detected in the $r$-band at any phase. 

    \item We obtained multi-epoch NIR spectra from SOAR/TSpec and VLT/KMOS, covering a wavelength range between 0.8 and 2.45~$\mu$m. SPICY~30021 has classical FUor spectra with key absorption features such as CO bandheads, water vapour and a blueshifted He~{\sc i} line. The EWs of Na~{\sc i} and Ca~{\sc i} are consistent with FUors, in contrast to late-type stars. However, the Pa $\beta$ absorption line has a symmetric shape and is not blueshifted, which could come from the accretion disk instead of a wind origin. By matching the $H$-bandpass continuum with FU~Ori, we estimated the extinction towards SPICY~30021 is $A_V = 12.5$ mag.

    \item We measured the distance to SPICY~30021 by combining the RV derived from the CO bandhead absorption with PMs from VIRAC2 and the Galactic spiral arm geometry, yielding a consistent estimate of $d = 1.7\pm 0.4$~kpc. {Based on the extinction, distance and PMS evolution models, we derived SPICY~30021 as a 0.9~$M_\odot$ star with 0.5~Myr age (distance-dependent)}. 
    
    \item Comparing with other bona fide FUors, SPICY~30021 sits at the {mid-to-low} end of eruptive amplitude in the optical and NIR, although its $W2$-amplitude is typical among others. We estimated an accretion luminosity of $L_{\rm acc} = {30}~\rm L_\odot$, which is common among FUors.  

\end{itemize}

SPICY~30021 is a new member of the bona fide FUor group, discovered by IR photometric surveys. Although {optically obscured}, the IR photometry and spectroscopic data revealed its photospheric and eruptive nature. By comparing the NIR and MIR light curves, we identified pre-outburst brightening in the MIR light curve. However, no spectrum has been observed during the rising stage of this event, which would be key to understanding the physical mechanisms of episodic mass accretion. 



\begin{acknowledgements}
We thank the anonymous referee for their constructive comments and helpful suggestions, which have improved the clarity and quality of the manuscript. We thank Dr Michael Connelly for providing the IRTF spectrum of FU Ori. 
Based on observations taken at ESO Paranal Observatory, for the KMOS VVVX-GalCen Spectroscopic Public 
Survey, ESO Programme ID: 116.29EF. The data are published in the ESO Science Archive Facility and are 
identified by the DOI: \url{https://doi.eso.org/10.18727/archive/38}. 

 This work is supported by the China-Chile Joint Research Fund (CCJRF No.2301) and the Chinese Academy of Sciences South America Center for Astronomy (CASSACA) Key Research Project E52H540301. ZG is supported by FONDECYT Iniciación project 11260176.  ZG and CM are funded by the project ALMA-ANID 31240014.
 
  JB and RK thank the support from FONDECYT Regular Project No. 1240249 and 1261142. VE acknowledges support from the Ministry of Science and Higher Education of the Russian Federation (State contract FENW-2026-0028). 

FN thanks CNPq for support through process 303093/2025-0. PW acknowledges support from ANID -- Millennium Science Initiative Program -- Center Code NCN2024\_001.
D.M. and C.C. are supported by the ANID BASAL Center for Astrophysics and Associated Tecnologies (CATA) project FB210003. MGN thanks the European Union - NextGenerationEU, M4C2 1.2 CUP C83C25000450006. 
Tracing the staR and plAnet formation in different Circumstellar Environments (TRACE)
"This work was funded by the European Union - NextGenerationEU, M4C2 1.2 CUP C83C25000450006”

MGN acknowledge financial support from the project "Space Observations of Proto-planetary Accretion Disc Evolution (SPADE)” in the framework of the "PROgramma di RIcerca Spaziale di base (PRORIS)", financed by the Italian Ministry of University and Research (Ministero dell’Università e della Ricerca; MUR).

M.G. and S.F. acknowledge the support from Fondecyt Regular 1240755. I.P. acknowledges support from ANID BECAS/DOCTORADO NACIONAL 21230761. C. O. Obasi is supported by the Postdoctoral Talent Attraction Competition for Research Centres and Institutes of Universidad Andrés Bello (UNAB) 2025, project No. DI-07-25/ATP. J.G.F-T gratefully acknowledges the support provided by ANID Fondecyt Regular No. 1260371.
J.A.-G. is supported by ANID Fondecyt Regular No. 1201490.

This publication makes use of data products from the Near-Earth Object Wide-field IR Survey Explorer (NEOWISE), which is a joint project of the Jet Propulsion Laboratory/California Institute of Technology and the University of Arizona. NEOWISE is funded by the National Aeronautics and Space Administration. This research has made use of the NASA/IPAC IR Science Archive, which is funded by the National Aeronautics and Space Administration and operated by the California Institute of Technology. Based on observations obtained at the Southern Astrophysical Research (SOAR) telescope, which is a joint project of the Minist\'{e}rio da Ci\^{e}ncia, Tecnologia e Inova\c{c}\~{o}es (MCTI/LNA) do Brasil, the U.S. National Science Foundation NOIRLab, the University of North Carolina at Chapel Hill (UNC), and Michigan State University (MSU). This work makes use of observations from the Las Cumbres Observatory global telescope network.

This research has made use of the SVO Filter Profile Service "Carlos Rodrigo", funded by MCIN/AEI/10.13039/501100011033/ through grant PID2020-112949GB-I00.

This work makes use of observations from the Las Cumbres Observatory global telescope network.

Based on observations obtained through the Astronomical Event Observatory Network (AEON), a joint endeavor of the Las Cumbres Observatory and of NSF NOIRLab, which is managed by the Association of Universities for Research in Astronomy (AURA) under a cooperative agreement with the U.S. National Science Foundation.

The authors acknowledge the use of artificial intelligence language models (ChatGPT and Grammarly) for assistance with text editing.
\end{acknowledgements}

\bibliographystyle{aa}
\bibliography{reference}

@ARTICLE{Portnoi2026,
       author = {{Portnoi}, Evan R. and {Hillenbrand}, Lynne A. and {Carvalho}, Adolfo S.},
        title = "{Quantitative Spectroscopic Diagnostics for FU Orionis-type Young Stellar Objects}",
      journal = {\apj},
         year = 2026,
        month = jul,
       volume = {1005},
       number = {1},
          eid = {57},
        pages = {57},
          doi = {10.3847/1538-4357/ae6dbc},
       adsurl = {https://ui.adsabs.harvard.edu/abs/2026ApJ..1005...57P}
}

@ARTICLE{Zucker2025,
       author = {{Zucker}, Catherine and {Saydjari}, Andrew K. and {Speagle}, Joshua S. and {Schlafly}, Edward F. and {Green}, Gregory M. and {Benjamin}, Robert and {Peek}, Joshua and {Edenhofer}, Gordian and {Goodman}, Alyssa and {Kuhn}, Michael A. and {Finkbeiner}, Douglas P.},
        title = "{A Deep, High-angular-resolution 3D Dust Map of the Southern Galactic Plane}",
      journal = {\apj},
         year = 2025,
        month = oct,
       volume = {992},
       number = {1},
          eid = {39},
        pages = {39},
          doi = {10.3847/1538-4357/adfbe6},
archivePrefix = {arXiv},
       eprint = {2503.02657},
 primaryClass = {astro-ph.GA},
       adsurl = {https://ui.adsabs.harvard.edu/abs/2025ApJ...992...39Z}
}

@ARTICLE{Movsessian2025,
       author = {{Movsessian}, T.~A. and {Bally}, J. and {Magakian}, T. Yu and {Moiseev}, A.~V.},
        title = "{Complex investigations of an active star formation region in the southern part of Mon R2}",
      journal = {\mnras},
         year = 2025,
        month = feb,
       volume = {536},
       number = {4},
        pages = {3251-3261},
          doi = {10.1093/mnras/stae2791},
archivePrefix = {arXiv},
       eprint = {2412.11268},
 primaryClass = {astro-ph.SR},
       adsurl = {https://ui.adsabs.harvard.edu/abs/2025MNRAS.536.3251M}
}

@ARTICLE{Movsessian2021,
       author = {{Movsessian}, T.~A. and {Magakian}, T. Yu and {Dodonov}, S.~N.},
        title = "{New Herbig-Haro objects and outflows in the Mon R1 association}",
      journal = {\mnras},
         year = 2021,
        month = jan,
       volume = {500},
       number = {2},
        pages = {2440-2450},
          doi = {10.1093/mnras/staa3302},
archivePrefix = {arXiv},
       eprint = {2005.04257},
 primaryClass = {astro-ph.SR},
       adsurl = {https://ui.adsabs.harvard.edu/abs/2021MNRAS.500.2440M}
}

@ARTICLE{Magakian2026,
       author = {{Magakian}, T. Yu. and {Movsessian}, T.~A. and {Moiseev}, A.~V. and {Molyarova}, T.~S. and {Uklein}, R.~I.},
        title = "{Optical spectroscopy of the IR source CPM 19 and surrounding objects}",
      journal = {Astrophysics},
         year = 2026,
        month = aug,
          doi = {10.1007/s10511-026-09963-9},
archivePrefix = {arXiv},
       eprint = {2603.22462},
 primaryClass = {astro-ph.SR},
       adsurl = {https://ui.adsabs.harvard.edu/abs/2026Ap....tmp...53M}
}

@ARTICLE{Roman2025,
       author = {{Galactic Plane Survey Definition Committee}, Roman},
        title = "{Roman Galactic Plane Survey Definition Committee Report}",
      journal = {arXiv e-prints},
         year = 2025,
        month = nov,
          eid = {arXiv:2511.07494},
        pages = {arXiv:2511.07494},
          doi = {10.48550/arXiv.2511.07494},
archivePrefix = {arXiv},
       eprint = {2511.07494},
 primaryClass = {astro-ph.GA},
       adsurl = {https://ui.adsabs.harvard.edu/abs/2025arXiv251107494G}
}

@ARTICLE{Contreras2025a,
       author = {{Contreras Pe{\~n}a}, Carlos and {Lee}, Jeong-Eun and {Lee}, Ho-Gyu and {Herczeg}, Gregory and {Johnstone}, Doug and {Liu}, Hanpu and {Lucas}, Philip W. and {Guo}, Zhen and {Kuhn}, Michael A. and {Smith}, Leigh C. and {Ashraf}, Mizna and {Jose}, Jessy and {Yoon}, Sung-Yong and {Yoon}, Sung-Chul},
        title = "{``Oh FUors, Where Art Thou?'': A Search for Long-lasting Young Stellar Object Outbursts Hiding in Infrared Surveys}",
      journal = {\apj},
         year = 2025,
        month = jul,
       volume = {987},
       number = {1},
          eid = {23},
        pages = {23},
          doi = {10.3847/1538-4357/add25f},
archivePrefix = {arXiv},
       eprint = {2504.21237},
 primaryClass = {astro-ph.SR},
       adsurl = {https://ui.adsabs.harvard.edu/abs/2025ApJ...987...23C}
}

@ARTICLE{Li2024,
       author = {{Li}, Jiaxun and {Wang}, Tinggui},
        title = "{A population of mid-infrared large-amplitude variable young stellar objects from unTimely}",
      journal = {\mnras},
         year = 2024,
        month = aug,
       volume = {532},
       number = {2},
        pages = {2683-2704},
          doi = {10.1093/mnras/stae1601},
archivePrefix = {arXiv},
       eprint = {2406.18303},
 primaryClass = {astro-ph.SR},
       adsurl = {https://ui.adsabs.harvard.edu/abs/2024MNRAS.532.2683L}
}

@ARTICLE{Bae2014,
       author = {{Bae}, Jaehan and {Hartmann}, Lee and {Zhu}, Zhaohuan and {Nelson}, Richard P.},
        title = "{Accretion Outbursts in Self-gravitating Protoplanetary Disks}",
      journal = {\apj},
         year = 2014,
        month = nov,
       volume = {795},
       number = {1},
          eid = {61},
        pages = {61},
          doi = {10.1088/0004-637X/795/1/61},
archivePrefix = {arXiv},
       eprint = {1409.3891},
 primaryClass = {astro-ph.SR},
       adsurl = {https://ui.adsabs.harvard.edu/abs/2014ApJ...795...61B}
}

@ARTICLE{Carvalho2024b,
       author = {{Carvalho}, Adolfo and {Hillenbrand}, Lynne and {Seebeck}, Jerome and {Covey}, Kevin},
        title = "{An Expanding Accretion Disk and a Warm Disk Wind as Seen in the Spectral Evolution of HBC 722}",
      journal = {\apj},
         year = 2024,
        month = aug,
       volume = {971},
       number = {1},
          eid = {44},
        pages = {44},
          doi = {10.3847/1538-4357/ad5286},
archivePrefix = {arXiv},
       eprint = {2405.20251},
 primaryClass = {astro-ph.SR},
       adsurl = {https://ui.adsabs.harvard.edu/abs/2024ApJ...971...44C}
}

@ARTICLE{Guo2026,
       author = {{Guo}, Zhen and {Osses}, J. and {Fermiano}, V. and {Zhou}, Yuting and {Fang}, Min and {Herczeg}, G. and {Carvalho}, A. and {Elbakyan}, V. and {Hillenbrand}, L. and {Wang}, Mutian and {Liu}, Hanpu and {Liu}, Yao and {Brice{\~n}o}, C. and {Singh}, K. and {Ivanov}, V.~D. and {Ninan}, J. and {Giannini}, T. and {Aliaga}, A. and {Morris}, C. and {Montesinos}, M. and {Zhao}, He and {Contreras Pe{\~n}a}, C. and {Jose}, J. and {Chand}, T. and {Chen}, Wen-Ping and {Wang}, Wei-Hao and {Huang}, Yang and {Lopez}, C. and {Fernandez-Schlosser}, P. and {Correa-Herrera}, D. and {Kurtev}, R. and {Rodriguez}, V. and {Lizana-Vidal}, C. and {Borissova}, J. and {Kuhn}, M. and {Saito}, R.~K. and {Yadav}, Ram K.},
        title = "{Hot H$_{2}$O emission during an outburst in a classical T Tauri star}",
      journal = {\aap},
         year = 2026,
        month = oct,
       volume = {714},
          eid = {A98},
        pages = {A98},
          doi = {10.1051/0004-6361/202661691},
       adsurl = {https://ui.adsabs.harvard.edu/abs/2026A&A...714A..98G}
}

@ARTICLE{Contreras2026,
       author = {{Contreras Pe{\~n}a}, Carlos and {Lee}, Jeong-Eun and {Lucas}, Philip W. and {Herczeg}, Gregory and {Johnstone}, Doug and {Guo}, Zhen and {Lee}, Ho-Gyu and {Kim}, Hwan-Ki and {Jose}, Jessy and {Ashraf}, Mizna and {Morris}, Calum},
        title = "{The long-term outburst(s) of GPSV16: from an intermediate to a FUor classification}",
      journal = {\mnras},
         year = 2026,
        month = jun,
       volume = {549},
       number = {2},
          eid = {stag807},
        pages = {stag807},
          doi = {10.1093/mnras/stag807},
archivePrefix = {arXiv},
       eprint = {2604.24402},
 primaryClass = {astro-ph.SR},
       adsurl = {https://ui.adsabs.harvard.edu/abs/2026MNRAS.549ag807C}
}

@dataset{Cutri2003,
       author = {{Cutri}, R.~M. and {Skrutskie}, M.~F. and {van Dyk}, S. and {Beichman}, C.~A. and {Carpenter}, J.~M. and {Chester}, T. and {Cambresy}, L. and {Evans}, T. and {Fowler}, J. and {Gizis}, J. and {Howard}, E. and {Huchra}, J. and {Jarrett}, T. and {Kopan}, E.~L. and {Kirkpatrick}, J.~D. and {Light}, R.~M. and {Marsh}, K.~A. and {McCallon}, H. and {Schneider}, S. and {Stiening}, R. and {Sykes}, M. and {Weinberg}, M. and {Wheaton}, W.~A. and {Wheelock}, S. and {Zacarias}, N.},
        title = "{VizieR Online Data Catalog: 2MASS All-Sky Catalog of Point Sources (Cutri+ 2003)}",
 howpublished = {VizieR On-line Data Catalog: II/246.  Originally published in: University of Massachusetts and Infrared Processing and Analysis Center, (IPAC/California Institute of Technology) (2003)},
         year = 2003,
        month = jun,
          eid = {II/246},
       adsurl = {https://ui.adsabs.harvard.edu/abs/2003yCat.2246....0C}
}

@ARTICLE{Siwak2025,
       author = {{Siwak}, M. and {K{\'o}sp{\'a}l}, {\'A}. and {{\'A}brah{\'a}m}, P. and {Marton}, G. and {Zieli{\'n}ski}, P. and {Gromadzki}, M. and {Wyrzykowski}, {\L}. and {Nagy}, Z. and {Szil{\'a}gyi}, M. and {Potter}, S.~B. and {Sefako}, R. and {Worters}, H.~L. and {Buckley}, D.~A.~H. and {Giannini}, T. and {Fiorellino}, E. and {Cruz-S{\'a}enz de Miera}, F. and {Kun}, M. and {Szab{\'o}}, Zs. M. and {Lucas}, P.~W. and {Krzesi{\'n}ski}, J. and {Zakrzewski}, B. and {Og{\l}oza}, W. and {P{\'a}l}, A. and {Cseh}, B. and {Horti-D{\'a}vid}, {\'A}. and {Jo{\'o}}, A. and {Kalup}, Cs. and {Kriskovics}, L. and {S{\'o}dor}, {\'A}. and {Szak{\'a}ts}, R. and {Vink{\'o}}, J.},
        title = "{Gaia20bdk ─ New FU Ori-type star in the Sh 2-301 star-forming region}",
      journal = {\aap},
         year = 2025,
        month = mar,
       volume = {695},
          eid = {A130},
        pages = {A130},
          doi = {10.1051/0004-6361/202451061},
archivePrefix = {arXiv},
       eprint = {2412.07697},
 primaryClass = {astro-ph.SR},
       adsurl = {https://ui.adsabs.harvard.edu/abs/2025A&A...695A.130S}
}

@ARTICLE{Hillenbrand2026,
       author = {{Hillenbrand}, Lynne A. and {De}, Kishalay and {Carvalho}, Adolfo S. and {Stern}, Dan and {Portnoi}, Evan and {Earley}, Nicholas},
        title = "{V7995 Sgr: A New FU Orionis Accretion Outburst Near NGC 6589/6590}",
      journal = {arXiv e-prints},
         year = 2026,
        month = jun,
          eid = {arXiv:2606.20350},
        pages = {arXiv:2606.20350},
          doi = {10.48550/arXiv.2606.20350},
archivePrefix = {arXiv},
       eprint = {2606.20350},
 primaryClass = {astro-ph.SR},
       adsurl = {https://ui.adsabs.harvard.edu/abs/2026arXiv260620350H}
}

@ARTICLE{Elbakyan2025,
       author = {{Elbakyan}, Vardan G. and {Wehner}, Dennis and {Kuiper}, Rolf and {Nayakshin}, Sergei and {Caratti o Garatti}, Alessio and {Guo}, Zhen},
        title = "{Episodic accretion in high-mass star formation: An analysis of thermal instability for axially symmetric disks}",
      journal = {\aap},
         year = 2025,
        month = sep,
       volume = {701},
          eid = {A91},
        pages = {A91},
          doi = {10.1051/0004-6361/202555559},
archivePrefix = {arXiv},
       eprint = {2507.20781},
 primaryClass = {astro-ph.SR},
       adsurl = {https://ui.adsabs.harvard.edu/abs/2025A&A...701A..91E}
}

@ARTICLE{Chen2019,
       author = {{Chen}, B.-Q. and {Huang}, Y. and {Yuan}, H.-B. and {Wang}, C. and {Fan}, D.-W. and {Xiang}, M.-S. and {Zhang}, H.-W. and {Tian}, Z.-J. and {Liu}, X.-W.},
        title = "{Three-dimensional interstellar dust reddening maps of the Galactic plane}",
      journal = {\mnras},
         year = 2019,
        month = mar,
       volume = {483},
       number = {4},
        pages = {4277-4289},
          doi = {10.1093/mnras/sty3341},
archivePrefix = {arXiv},
       eprint = {1807.02241},
 primaryClass = {astro-ph.GA},
       adsurl = {https://ui.adsabs.harvard.edu/abs/2019MNRAS.483.4277C}
}

@ARTICLE{Zhang2026,
       author = {{Zhang}, Baisong and {Chen}, Bingqiu and {Fan}, Dongwei and {Yuan}, Haibo and {Chen}, Pinjian and {Guo}, Helong and {Zhang}, Lin and {Sun}, Baokun and {Zou}, Xingzhu and {Zhang}, Lunwei and {Cao}, Yanan and {Ding}, Longfei},
        title = "{GSED: The Galactic Stellar Extinction Database}",
      journal = {arXiv e-prints},
         year = 2026,
        month = jun,
          eid = {arXiv:2606.31374},
        pages = {arXiv:2606.31374},
archivePrefix = {arXiv},
       eprint = {2606.31374},
 primaryClass = {astro-ph.GA},
       adsurl = {https://ui.adsabs.harvard.edu/abs/2026arXiv260631374Z}
}

@INPROCEEDINGS{Autry2003,
       author = {{Autry}, Roy G. and {Probst}, Ronald G. and {Starr}, Barry M. and {Abdel-Gawad}, Khairy M. and {Blakley}, Rick D. and {Daly}, Philip N. and {Dominguez}, Ruben and {Hileman}, Edward A. and {Liang}, Ming and {Pearson}, Earl T. and {Shaw}, Richard A. and {Tody}, Doug},
        title = "{NEWFIRM: the widefield IR imager for NOAO 4-m telescopes}",
    booktitle = {Instrument Design and Performance for Optical/Infrared Ground-based Telescopes},
         year = 2003,
       editor = {{Iye}, Masanori and {Moorwood}, Alan F.~M.},
       series = {Society of Photo-Optical Instrumentation Engineers (SPIE) Conference Series},
       volume = {4841},
        month = mar,
        pages = {525-539},
          doi = {10.1117/12.460419},
       adsurl = {https://ui.adsabs.harvard.edu/abs/2003SPIE.4841..525A}
}

@ARTICLE{Alerce2021,
       author = {{F{\"o}rster}, F. and {Cabrera-Vives}, G. and {Castillo-Navarrete}, E. and {Est{\'e}vez}, P.~A. and {S{\'a}nchez-S{\'a}ez}, P. and {Arredondo}, J. and {Bauer}, F.~E. and {Carrasco-Davis}, R. and {Catelan}, M. and {Elorrieta}, F. and {Eyheramendy}, S. and {Huijse}, P. and {Pignata}, G. and {Reyes}, E. and {Reyes}, I. and {Rodr{\'\i}guez-Mancini}, D. and {Ruz-Mieres}, D. and {Valenzuela}, C. and {{\'A}lvarez-Maldonado}, I. and {Astorga}, N. and {Borissova}, J. and {Clocchiatti}, A. and {De Cicco}, D. and {Donoso-Oliva}, C. and {Hern{\'a}ndez-Garc{\'\i}a}, L. and {Graham}, M.~J. and {Jord{\'a}n}, A. and {Kurtev}, R. and {Mahabal}, A. and {Maureira}, J.~C. and {Mu{\~n}oz-Arancibia}, A. and {Molina-Ferreiro}, R. and {Moya}, A. and {Palma}, W. and {P{\'e}rez-Carrasco}, M. and {Protopapas}, P. and {Romero}, M. and {Sabatini-Gacitua}, L. and {S{\'a}nchez}, A. and {San Mart{\'\i}n}, J. and {Sep{\'u}lveda-Cobo}, C. and {Vera}, E. and {Vergara}, J.~R.},
        title = "{The Automatic Learning for the Rapid Classification of Events (ALeRCE) Alert Broker}",
      journal = {\aj},
         year = 2021,
        month = may,
       volume = {161},
       number = {5},
          eid = {242},
        pages = {242},
          doi = {10.3847/1538-3881/abe9bc},
archivePrefix = {arXiv},
       eprint = {2008.03303},
 primaryClass = {astro-ph.IM},
       adsurl = {https://ui.adsabs.harvard.edu/abs/2021AJ....161..242F}
}

@ARTICLE{Morris2025,
       author = {{Morris}, C. and {Guo}, Z. and {Lucas}, P.~W. and {Miller}, N. and {Contreras Pe{\~n}a}, C. and {Kuhn}, M.~A.},
        title = "{Eruptive YSOs in Cygnus-X: a mid-infrared variability study with NEOWISE and SPICY}",
      journal = {\mnras},
         year = 2025,
        month = mar,
       volume = {537},
       number = {3},
        pages = {2763-2781},
          doi = {10.1093/mnras/staf079},
archivePrefix = {arXiv},
       eprint = {2501.03929},
 primaryClass = {astro-ph.SR},
       adsurl = {https://ui.adsabs.harvard.edu/abs/2025MNRAS.537.2763M}
}

@INPROCEEDINGS{Schlawin2014,
       author = {{Schlawin}, E. and {Herter}, T.~L. and {Henderson}, C. and {Wilson}, J.~C. and {Probst}, R. and {Sprayberry}, D. and {Bonati}, M. and {Schurter}, P. and {James}, D. and {Warner}, M. and {Tighe}, R. and {Adams}, J.~D. and {Mart{\'\i}nez}, M.},
        title = "{Design updates and status of the fourth generation TripleSpec spectrograph}",
    booktitle = {Ground-based and Airborne Instrumentation for Astronomy V},
         year = 2014,
       editor = {{Ramsay}, Suzanne K. and {McLean}, Ian S. and {Takami}, Hideki},
       series = {Society of Photo-Optical Instrumentation Engineers (SPIE) Conference Series},
       volume = {9147},
        month = aug,
          eid = {91472H},
        pages = {91472H},
          doi = {10.1117/12.2055233},
       adsurl = {https://ui.adsabs.harvard.edu/abs/2014SPIE.9147E..2HS}
}

@ARTICLE{Kuhn2021,
       author = {{Kuhn}, Michael A. and {de Souza}, Rafael S. and {Krone-Martins}, Alberto and {Castro-Ginard}, Alfred and {Ishida}, Emille E.~O. and {Povich}, Matthew S. and {Hillenbrand}, Lynne A. and {COIN Collaboration}},
        title = "{SPICY: The Spitzer/IRAC Candidate YSO Catalog for the Inner Galactic Midplane}",
      journal = {\apjs},
         year = 2021,
        month = jun,
       volume = {254},
       number = {2},
          eid = {33},
        pages = {33},
          doi = {10.3847/1538-4365/abe465},
archivePrefix = {arXiv},
       eprint = {2011.12961},
 primaryClass = {astro-ph.GA},
       adsurl = {https://ui.adsabs.harvard.edu/abs/2021ApJS..254...33K}
}

@ARTICLE{Guo2021,
       author = {{Guo}, Zhen and {Lucas}, P.~W. and {Contreras Pe{\~n}a}, C. and {Smith}, L.~C. and {Morris}, C. and {Kurtev}, R.~G. and {Borissova}, J. and {Alonso-Garc{\'\i}a}, J. and {Minniti}, D. and {Chen{\'e}}, A. -N. and {Kumar}, M.~S.~N. and {Caratti o Garatti}, A. and {Froebrich}, D. and {Stimson}, W.~H.},
        title = "{Analysis of physical processes in eruptive YSOs with near-infrared spectra and multiwavelength light curves}",
      journal = {\mnras},
         year = 2021,
        month = jun,
       volume = {504},
       number = {1},
        pages = {830-856},
          doi = {10.1093/mnras/stab882},
archivePrefix = {arXiv},
       eprint = {2103.13335},
 primaryClass = {astro-ph.SR},
       adsurl = {https://ui.adsabs.harvard.edu/abs/2021MNRAS.504..830G}
}

@ARTICLE{Bensby2013,
       author = {{Bensby}, T. and {Yee}, J.~C. and {Feltzing}, S. and {Johnson}, J.~A. and
         {Gould}, A. and {Cohen}, J.~G. and {Asplund}, M. and
         {Mel{\'e}ndez}, J. and {Lucatello}, S. and {Han}, C. and
         {Thompson}, I. and {Gal-Yam}, A. and {Udalski}, A. and
         {Bennett}, D.~P. and {Bond}, I.~A. and {Kohei}, W. and {Sumi}, T. and
         {Suzuki}, D. and {Suzuki}, K. and {Takino}, S. and {Tristram}, P. and
         {Yamai}, N. and {Yonehara}, A.},
        title = "{Chemical evolution of the Galactic bulge as traced by microlensed dwarf and subgiant stars. V. Evidence for a wide age distribution and a complex MDF}",
      journal = {\aap},
         year = 2013,
        month = jan,
       volume = {549},
          eid = {A147},
        pages = {A147},
          doi = {10.1051/0004-6361/201220678},
archivePrefix = {arXiv},
       eprint = {1211.6848},
 primaryClass = {astro-ph.GA},
       adsurl = {https://ui.adsabs.harvard.edu/abs/2013A&A...549A.147B}
}

@ARTICLE{Lodato2004,
       author = {{Lodato}, G. and {Clarke}, C.~J.},
        title = "{Massive planets in FU Orionis discs: implications for thermal instability models}",
      journal = {\mnras},
         year = 2004,
        month = sep,
       volume = {353},
       number = {3},
        pages = {841-852},
          doi = {10.1111/j.1365-2966.2004.08112.x},
archivePrefix = {arXiv},
       eprint = {astro-ph/0406264},
 primaryClass = {astro-ph},
       adsurl = {https://ui.adsabs.harvard.edu/abs/2004MNRAS.353..841L}
}

@ARTICLE{Mainzer2014,
       author = {{Mainzer}, A. and {Bauer}, J. and {Cutri}, R.~M. and {Grav}, T. and
         {Masiero}, J. and {Beck}, R. and {Clarkson}, P. and {Conrow}, T. and
         {Dailey}, J. and {Eisenhardt}, P. and {Fabinsky}, B. and
         {Fajardo-Acosta}, S. and {Fowler}, J. and {Gelino}, C. and
         {Grillmair}, C. and {Heinrichsen}, I. and {Kendall}, M. and
         {Kirkpatrick}, J. Davy and {Liu}, F. and {Masci}, F. and
         {McCallon}, H. and {Nugent}, C.~R. and {Papin}, M. and {Rice}, E. and
         {Royer}, D. and {Ryan}, T. and {Sevilla}, P. and {Sonnett}, S. and
         {Stevenson}, R. and {Thompson}, D.~B. and {Wheelock}, S. and
         {Wiemer}, D. and {Wittman}, M. and {Wright}, E. and {Yan}, L.},
        title = "{Initial Performance of the NEOWISE Reactivation Mission}",
      journal = {\apj},
         year = 2014,
        month = sep,
       volume = {792},
       number = {1},
          eid = {30},
        pages = {30},
          doi = {10.1088/0004-637X/792/1/30},
archivePrefix = {arXiv},
       eprint = {1406.6025},
 primaryClass = {astro-ph.EP},
       adsurl = {https://ui.adsabs.harvard.edu/abs/2014ApJ...792...30M}
}

@ARTICLE{Reid2014,
       author = {{Reid}, M.~J. and {Menten}, K.~M. and {Brunthaler}, A. and
         {Zheng}, X.~W. and {Dame}, T.~M. and {Xu}, Y. and {Wu}, Y. and
         {Zhang}, B. and {Sanna}, A. and {Sato}, M. and {Hachisuka}, K. and
         {Choi}, Y.~K. and {Immer}, K. and {Moscadelli}, L. and
         {Rygl}, K.~L.~J. and {Bartkiewicz}, A.},
        title = "{Trigonometric Parallaxes of High Mass Star Forming Regions: The Structure and Kinematics of the Milky Way}",
      journal = {\apj},
         year = 2014,
        month = mar,
       volume = {783},
       number = {2},
          eid = {130},
        pages = {130},
          doi = {10.1088/0004-637X/783/2/130},
archivePrefix = {arXiv},
       eprint = {1401.5377},
 primaryClass = {astro-ph.GA},
       adsurl = {https://ui.adsabs.harvard.edu/abs/2014ApJ...783..130R}
}

@ARTICLE{Wenger2018,
       author = {{Wenger}, Trey V. and {Balser}, Dana S. and {Anderson}, L.~D. and
         {Bania}, T.~M.},
        title = "{Kinematic Distances: A Monte Carlo Method}",
      journal = {\apj},
         year = 2018,
        month = mar,
       volume = {856},
       number = {1},
          eid = {52},
        pages = {52},
          doi = {10.3847/1538-4357/aaaec8},
archivePrefix = {arXiv},
       eprint = {1802.04203},
 primaryClass = {astro-ph.GA},
       adsurl = {https://ui.adsabs.harvard.edu/abs/2018ApJ...856...52W}
}

@ARTICLE{Armitage2001,
       author = {{Armitage}, Philip J. and {Livio}, Mario and {Pringle}, J.~E.},
        title = "{Episodic accretion in magnetically layered protoplanetary discs}",
      journal = {\mnras},
         year = 2001,
        month = jun,
       volume = {324},
       number = {3},
        pages = {705-711},
          doi = {10.1046/j.1365-8711.2001.04356.x},
archivePrefix = {arXiv},
       eprint = {astro-ph/0101253},
 primaryClass = {astro-ph},
       adsurl = {https://ui.adsabs.harvard.edu/abs/2001MNRAS.324..705A}
}

@ARTICLE{Cuello2019,
       author = {{Cuello}, Nicol{\'a}s and {Dipierro}, Giovanni and {Mentiplay}, Daniel and {Price}, Daniel J. and {Pinte}, Christophe and {Cuadra}, Jorge and {Laibe}, Guillaume and {M{\'e}nard}, Fran{\c{c}}ois and {Poblete}, Pedro P. and {Montesinos}, Mat{\'\i}as},
        title = "{Flybys in protoplanetary discs: I. Gas and dust dynamics}",
      journal = {\mnras},
         year = 2019,
        month = mar,
       volume = {483},
       number = {3},
        pages = {4114-4139},
          doi = {10.1093/mnras/sty3325},
archivePrefix = {arXiv},
       eprint = {1812.00961},
 primaryClass = {astro-ph.EP},
       adsurl = {https://ui.adsabs.harvard.edu/abs/2019MNRAS.483.4114C}
}

@ARTICLE{Montesinos2026,
       author = {{Montesinos}, Mat{\'\i}as and {Nayakshin}, Sergei and {Elbakyan}, Vardan and {Guo}, Zhen and {Sucerquia}, Mario and {Bayo}, Amelia and {Zhu}, Zhaohuan},
        title = "{Observational signatures of planetary tidal disruption events around solar-mass stars}",
      journal = {\aap},
         year = 2026,
        month = mar,
       volume = {707},
          eid = {A222},
        pages = {A222},
          doi = {10.1051/0004-6361/202558389},
archivePrefix = {arXiv},
       eprint = {2601.12501},
 primaryClass = {astro-ph.EP},
       adsurl = {https://ui.adsabs.harvard.edu/abs/2026A&A...707A.222M}
}

@ARTICLE{Liu2022,
       author = {{Liu}, Hanpu and {Herczeg}, Gregory J. and {Johnstone}, Doug and {Contreras-Pe{\~n}a}, Carlos and {Lee}, Jeong-Eun and {Yang}, Haifeng and {Zhou}, Xingyu and {Yoon}, Sung-Yong and {Lee}, Ho-Gyu and {Kunitomo}, Masanobu and {Jose}, Jessy},
        title = "{Diagnosing FU Ori-like Sources: The Parameter Space of Viscously Heated Disks in the Optical and Near-infrared}",
      journal = {\apj},
         year = 2022,
        month = sep,
       volume = {936},
       number = {2},
          eid = {152},
        pages = {152},
          doi = {10.3847/1538-4357/ac84d2},
archivePrefix = {arXiv},
       eprint = {2207.13324},
 primaryClass = {astro-ph.SR},
       adsurl = {https://ui.adsabs.harvard.edu/abs/2022ApJ...936..152L}
}

@INPROCEEDINGS{Fischer2023,
       author = {{Fischer}, W.~J. and {Hillenbrand}, L.~A. and {Herczeg}, G.~J. and {Johnstone}, D. and {Kospal}, A. and {Dunham}, M.~M.},
        title = "{Accretion Variability as a Guide to Stellar Mass Assembly}",
    booktitle = {Protostars and Planets VII},
         year = 2023,
       editor = {{Inutsuka}, S. and {Aikawa}, Y. and {Muto}, T. and {Tomida}, K. and {Tamura}, M.},
       series = {Astronomical Society of the Pacific Conference Series},
       volume = {534},
        month = jul,
        pages = {355},
          doi = {10.48550/arXiv.2203.11257},
archivePrefix = {arXiv},
       eprint = {2203.11257},
 primaryClass = {astro-ph.SR},
       adsurl = {https://ui.adsabs.harvard.edu/abs/2023ASPC..534..355F}
}

@ARTICLE{Vorobyov2010,
       author = {{Vorobyov}, Eduard I. and {Basu}, Shantanu},
        title = "{The Burst Mode of Accretion and Disk Fragmentation in the Early Embedded Stages of Star Formation}",
      journal = {\apj},
         year = 2010,
        month = aug,
       volume = {719},
       number = {2},
        pages = {1896-1911},
          doi = {10.1088/0004-637X/719/2/1896},
archivePrefix = {arXiv},
       eprint = {1007.2993},
 primaryClass = {astro-ph.SR},
       adsurl = {https://ui.adsabs.harvard.edu/abs/2010ApJ...719.1896V}
}

@ARTICLE{Zhu2009b,
       author = {{Zhu}, Zhaohuan and {Hartmann}, Lee and {Gammie}, Charles},
        title = "{Nonsteady Accretion in Protostars}",
      journal = {\apj},
         year = 2009,
        month = apr,
       volume = {694},
       number = {2},
        pages = {1045-1055},
          doi = {10.1088/0004-637X/694/2/1045},
archivePrefix = {arXiv},
       eprint = {0811.1762},
 primaryClass = {astro-ph},
       adsurl = {https://ui.adsabs.harvard.edu/abs/2009ApJ...694.1045Z}
}

@ARTICLE{Zhu2009,
       author = {{Zhu}, Zhaohuan and {Espaillat}, Catherine and {Hinkle}, Kenneth and
         {Hernandez}, Jesus and {Hartmann}, Lee and {Calvet}, Nuria},
        title = "{The Differential Rotation of FU Ori}",
      journal = {\apjl},
         year = 2009,
        month = mar,
       volume = {694},
       number = {1},
        pages = {L64-L68},
          doi = {10.1088/0004-637X/694/1/L64},
archivePrefix = {arXiv},
       eprint = {0902.4469},
 primaryClass = {astro-ph.SR},
       adsurl = {https://ui.adsabs.harvard.edu/abs/2009ApJ...694L..64Z}
}

@ARTICLE{Smette2015,
       author = {{Smette}, A. and {Sana}, H. and {Noll}, S. and {Horst}, H. and
         {Kausch}, W. and {Kimeswenger}, S. and {Barden}, M. and {Szyszka}, C. and
         {Jones}, A.~M. and {Gallenne}, A. and {Vinther}, J. and
         {Ballester}, P. and {Taylor}, J.},
        title = "{Molecfit: A general tool for telluric absorption correction. I. Method and application to ESO instruments}",
      journal = {\aap},
         year = 2015,
        month = apr,
       volume = {576},
          eid = {A77},
        pages = {A77},
          doi = {10.1051/0004-6361/201423932},
archivePrefix = {arXiv},
       eprint = {1501.07239},
 primaryClass = {astro-ph.IM},
       adsurl = {https://ui.adsabs.harvard.edu/abs/2015A&A...576A..77S}
}

@INPROCEEDINGS{Rodrigo2020,
       author = {{Rodrigo}, C. and {Solano}, E.},
        title = "{The SVO Filter Profile Service}",
    booktitle = {XIV.0 Scientific Meeting (virtual) of the Spanish Astronomical Society},
         year = 2020,
        month = jul,
          eid = {182},
        pages = {182},
       adsurl = {https://ui.adsabs.harvard.edu/abs/2020sea..confE.182R}
}

@ARTICLE{Guo2020,
       author = {{Guo}, Zhen and {Lucas}, P.~W. and {Contreras Pe{\~n}a}, C. and
         {Kurtev}, R.~G. and {Smith}, L.~C. and {Borissova}, J. and
         {Alonso-Garc{\'\i}a}, J. and {Minniti}, D. and {Caratti o Garatti}, A. and
         {Froebrich}, D.},
        title = "{Short- and long-term near-infrared spectroscopic variability of eruptive protostars from VVV}",
      journal = {\mnras},
         year = 2020,
        month = feb,
       volume = {492},
       number = {1},
        pages = {294-314},
          doi = {10.1093/mnras/stz3374},
archivePrefix = {arXiv},
       eprint = {1912.00729},
 primaryClass = {astro-ph.SR},
       adsurl = {https://ui.adsabs.harvard.edu/abs/2020MNRAS.492..294G}
}

@ARTICLE{Hillenbrand2018,
   author = {{Hillenbrand}, L.~A. and {Contreras Pe{\~n}a}, C. and {Morrell}, S. and 
  {Naylor}, T. and {Kuhn}, M.~A. and {Cutri}, R.~M. and {Rebull}, L.~M. and 
  {Hodgkin}, S. and {Froebrich}, D. and {Mainzer}, A.~K.},
    title = "{Gaia 17bpi: An FU Ori-type Outburst}",
  journal = {\apj},
archivePrefix = "arXiv",
   eprint = {1812.06640},
 primaryClass = "astro-ph.SR",
     year = 2018,
    month = dec,
   volume = 869,
      eid = {146},
    pages = {146},
      doi = {10.3847/1538-4357/aaf414},
   adsurl = {https://ui.adsabs.harvard.edu/abs/2018ApJ...869..146H}
}

@ARTICLE{Bell1994,
       author = {{Bell}, K.~R. and {Lin}, D.~N.~C.},
        title = "{Using FU Orionis Outbursts to Constrain Self-regulated Protostellar Disk Models}",
      journal = {\apj},
         year = 1994,
        month = jun,
       volume = {427},
        pages = {987},
          doi = {10.1086/174206},
archivePrefix = {arXiv},
       eprint = {astro-ph/9312015},
 primaryClass = {astro-ph},
       adsurl = {https://ui.adsabs.harvard.edu/abs/1994ApJ...427..987B}
}

@ARTICLE{Connelley2018,
   author = {{Connelley}, M.~S. and {Reipurth}, B.},
    title = "{A Near-infrared Spectroscopic Survey of FU Orionis Objects}",
  journal = {\apj},
archivePrefix = "arXiv",
   eprint = {1806.08880},
 primaryClass = "astro-ph.SR",
     year = 2018,
    month = jul,
   volume = 861,
      eid = {145},
    pages = {145},
      doi = {10.3847/1538-4357/aaba7b},
   adsurl = {http://adsabs.harvard.edu/abs/2018ApJ...861..145C}
}

@ARTICLE{Guo2025,
       author = {{Guo}, Z. and {Lucas}, P. and {Yurchenko}, S.~N. and {Kaminski}, T. and {Montesinos}, M. and {Nayakshin}, S. and {Elbakyan}, V. and {Osses}, J. and {Caratti o Garatti}, A. and {Zhao}, H. and {Kurtev}, R. and {Borissova}, J. and {Morris}, C. and {Minniti}, D. and {Alonso-Garc{\'\i}a}, J. and {Fermiano}, V. and {Saito}, R.~K. and {Miller}, N. and {Zsidi}, G. and {Muthu}, H.~D.~S. and {Brice{\~n}o}, C. and {Contreras Pe{\~n}a}, C. and {Lynas-Gray}, A.~E. and {Tennyson}, J. and {Wang}, L. and {Yu}, L. and {Benitez-Palacios}, D. and {Yang}, J. and {Kuhn}, M. and {Wang}, S.~X.},
        title = "{VVV-WIT-13: An eruptive young star with cool molecular features}",
      journal = {\aap},
         year = 2025,
        month = nov,
       volume = {703},
          eid = {A141},
        pages = {A141},
          doi = {10.1051/0004-6361/202556048},
archivePrefix = {arXiv},
       eprint = {2509.15334},
 primaryClass = {astro-ph.SR},
       adsurl = {https://ui.adsabs.harvard.edu/abs/2025A&A...703A.141G}
}

@ARTICLE{Contreras2025,
       author = {{Contreras Pe{\~n}a}, Carlos and {Lee}, Jeong-Eun and {Herczeg}, Gregory and {Johnstone}, Doug and {{\'A}brah{\'a}m}, P{\'e}ter and {Antoniucci}, Simone and {Audard}, Marc and {Ashraf}, Mizna and {Baek}, Giseon and {Garatti}, Alessio Caratti o. and {Carvalho}, Adolfo and {Cieza}, Lucas and {Cruz-Sa{\'e}nz de Miera}, Fernando and {Eisl{\"o}ffel}, Jochen and {Froebrich}, Dirk and {Giannini}, Teresa and {Green}, Joel and {Ghosh}, Arpan and {Guo}, Zhen and {Hillenbrand}, Lynne and {Hodapp}, Klaus and {Jheonn}, Hyunwook and {Jose}, Jessy and {Kim}, Young-Jun and {K{\'o}sp{\'a}l}, {\'A}gnes and {Lee}, Ho-Gyu and {Lucas}, Philip W. and {Magakian}, Tigran and {Nagy}, Zs{\'o}fia and {Naylor}, Tim and {Ninan}, Joe P. and {Peneva}, S. and {Reipurth}, Bo and {Scholz}, Alexander and {Semkov}, E. and {Sicilia-Aguilar}, Aurora and {Singh}, Koshvendra and {Siwak}, Michal and {Stecklum}, Bringfried and {Szab{\'o}}, Zs{\'o}fia Marianna and {Wolf}, Verena and {Yoon}, Sung-Yong},
        title = "{The Outbursting YSOs Catalogue (OYCAT)}",
      journal = {Journal of Korean Astronomical Society},
         year = 2025,
        month = sep,
       volume = {58},
        pages = {209-230},
          doi = {10.5303/JKAS.2025.58.2.209},
archivePrefix = {arXiv},
       eprint = {2509.24876},
 primaryClass = {astro-ph.SR},
       adsurl = {https://ui.adsabs.harvard.edu/abs/2025JKAS...58..209C}
}

@ARTICLE{Contreras2017b,
   author = {{Contreras Pe{\~n}a}, C. and {Lucas}, P.~W. and {Kurtev}, R. and 
  {Minniti}, D. and {Caratti o Garatti}, A. and {Marocco}, F. and 
  {Thompson}, M.~A. and {Froebrich}, D. and {Kumar}, M.~S.~N. and 
  {Stimson}, W. and {Navarro Molina}, C. and {Borissova}, J. and 
  {Gledhill}, T. and {Terzi}, R.},
    title = "{Infrared spectroscopy of eruptive variable protostars from VVV}",
  journal = {\mnras},
archivePrefix = "arXiv",
   eprint = {1602.06269},
 primaryClass = "astro-ph.SR",
     year = 2017,
    month = mar,
   volume = 465,
    pages = {3039-3100},
      doi = {10.1093/mnras/stw2802},
   adsurl = {http://adsabs.harvard.edu/abs/2017MNRAS.465.3039C}
}

@ARTICLE{Audard2014,
   author = {{Audard}, M. and {{\'A}brah{\'a}m}, P. and {Dunham}, M.~M. and 
  {Green}, J.~D. and {Grosso}, N. and {Hamaguchi}, K. and {Kastner}, J.~H. and 
  {K{\'o}sp{\'a}l}, {\'A}. and {Lodato}, G. and {Romanova}, M.~M. and 
  {Skinner}, S.~L. and {Vorobyov}, E.~I. and {Zhu}, Z.},
    title = "{Episodic Accretion in Young Stars}",
  journal = {Protostars and Planets VI},
archivePrefix = "arXiv",
   eprint = {1401.3368},
 primaryClass = "astro-ph.SR",
 publisher = {University of Arizonal Press, Chicago, IL},
     year = 2014,
    pages = {387-410},
      doi = {10.2458/azu_uapress_9780816531240-ch017},
   adsurl = {http://adsabs.harvard.edu/abs/2014prpl.conf..387A}
}

@ARTICLE{Cardelli1989,
   author = {{Cardelli}, J.~A. and {Clayton}, G.~C. and {Mathis}, J.~S.},
    title = "{The relationship between infrared, optical, and ultraviolet extinction}",
  journal = {\apj},
     year = 1989,
    month = oct,
   volume = 345,
    pages = {245-256},
      doi = {10.1086/167900},
   adsurl = {http://adsabs.harvard.edu/abs/1989ApJ...345..245C}
}

@ARTICLE{Allard2001,
   author = {{Allard}, F. and {Hauschildt}, P.~H. and {Alexander}, D.~R. and
  {Tamanai}, A. and {Schweitzer}, A.},
    title = "{The Limiting Effects of Dust in Brown Dwarf Model Atmospheres}",
  journal = {\apj},
   eprint = {astro-ph/0104256},
     year = 2001,
    month = jul,
   volume = 556,
    pages = {357-372},
      doi = {10.1086/321547},
   adsurl = {http://adsabs.harvard.edu/abs/2001ApJ...556..357A}
}

@ARTICLE{Cieza2016,
       author = {{Cieza}, Lucas A. and {Casassus}, Simon and {Tobin}, John and {Bos}, Steven P. and {Williams}, Jonathan P. and {Perez}, Sebastian and {Zhu}, Zhaohuan and {Caceres}, Claudio and {Canovas}, Hector and {Dunham}, Michael M. and {Hales}, Antonio and {Prieto}, Jose L. and {Principe}, David A. and {Schreiber}, Matthias R. and {Ruiz-Rodriguez}, Dary and {Zurlo}, Alice},
        title = "{Imaging the water snow-line during a protostellar outburst}",
      journal = {\nat},
         year = 2016,
        month = jul,
       volume = {535},
       number = {7611},
        pages = {258-261},
          doi = {10.1038/nature18612},
archivePrefix = {arXiv},
       eprint = {1607.03757},
 primaryClass = {astro-ph.SR},
       adsurl = {https://ui.adsabs.harvard.edu/abs/2016Natur.535..258C}
}

@ARTICLE{Smith2025,
       author = {{Smith}, Leigh C. and {Lucas}, Philip W. and {Koposov}, Sergey E. and {Gonzalez-Fernandez}, Carlos and {Alonso-Garc{\'\i}a}, Javier and {Minniti}, Dante and {Sanders}, Jason L. and {Bedin}, Luigi R. and {Belokurov}, Vasily and {Evans}, N. Wyn and {Hempel}, Maren and {Ivanov}, Valentin D. and {Kurtev}, Radostin G. and {Saito}, Roberto K.},
        title = "{VIRAC2: NIR astrometry and time series photometry for 500M+ stars from the VVV and VVVX surveys}",
      journal = {\mnras},
         year = 2025,
        month = feb,
       volume = {536},
       number = {4},
        pages = {3707-3738},
          doi = {10.1093/mnras/stae2797},
archivePrefix = {arXiv},
       eprint = {2501.06295},
 primaryClass = {astro-ph.GA},
       adsurl = {https://ui.adsabs.harvard.edu/abs/2025MNRAS.536.3707S}
}

@ARTICLE{Lee2019NatAs,
       author = {{Lee}, Jeong-Eun and {Lee}, Seokho and {Baek}, Giseon and {Aikawa}, Yuri and {Cieza}, Lucas and {Yoon}, Sung-Yong and {Herczeg}, Gregory and {Johnstone}, Doug and {Casassus}, Simon},
        title = "{The ice composition in the disk around V883 Ori revealed by its stellar outburst}",
      journal = {Nature Astronomy},
         year = 2019,
        month = feb,
       volume = {3},
        pages = {314-319},
          doi = {10.1038/s41550-018-0680-0},
archivePrefix = {arXiv},
       eprint = {1809.00353},
 primaryClass = {astro-ph.SR},
       adsurl = {https://ui.adsabs.harvard.edu/abs/2019NatAs...3..314L}
}

@ARTICLE{Das2026,
       author = {{Das}, Gautam and {Hillenbrand}, Lynne A. and {Carvalho}, Adolfo S.},
        title = "{A Parameterized YSO Accretion Disk Model with Increasing Accretion Rate: Predicted Outburst Lightcurves}",
      journal = {\apj},
         year = 2026,
        month = jun,
       volume = {1004},
       number = {2},
          eid = {201},
        pages = {201},
          doi = {10.3847/1538-4357/ae6e3b},
archivePrefix = {arXiv},
       eprint = {2605.19710},
 primaryClass = {astro-ph.SR},
       adsurl = {https://ui.adsabs.harvard.edu/abs/2026ApJ..1004..201D}
}

@ARTICLE{Cleaver2023,
       author = {{Cleaver}, Jacob and {Hartmann}, Lee and {Bae}, Jaehan},
        title = "{Magnetically activated accretion outbursts of pre-main-sequence discs}",
      journal = {\mnras},
         year = 2023,
        month = aug,
       volume = {523},
       number = {4},
        pages = {5522-5534},
          doi = {10.1093/mnras/stad1784},
archivePrefix = {arXiv},
       eprint = {2306.11098},
 primaryClass = {astro-ph.SR},
       adsurl = {https://ui.adsabs.harvard.edu/abs/2023MNRAS.523.5522C}
}

@ARTICLE{Hillenbrand2021,
       author = {{Hillenbrand}, Lynne A.},
        title = "{Spectral Features of YSO Outburst Gaia 21bty}",
      journal = {The Astronomer's Telegram},
         year = 2021,
        month = may,
       volume = {14590},
        pages = {1},
       adsurl = {https://ui.adsabs.harvard.edu/abs/2021ATel14590....1H}
}

@ARTICLE{Scholz2013,
   author = {{Scholz}, A. and {Froebrich}, D. and {Wood}, K.},
    title = "{A systematic survey for eruptive young stellar objects using mid-infrared photometry}",
  journal = {\mnras},
archivePrefix = "arXiv",
   eprint = {1301.3152},
 primaryClass = "astro-ph.SR",
     year = 2013,
    month = apr,
   volume = 430,
    pages = {2910-2922},
      doi = {10.1093/mnras/stt091},
   adsurl = {http://adsabs.harvard.edu/abs/2013MNRAS.430.2910S}
}

@ARTICLE{Hartmann1996,
   author = {{Hartmann}, L. and {Kenyon}, S.~J.},
    title = "{The FU Orionis Phenomenon}",
  journal = {\araa},
     year = 1996,
   volume = 34,
    pages = {207-240},
      doi = {10.1146/annurev.astro.34.1.207},
   adsurl = {http://adsabs.harvard.edu/abs/1996ARA%26A..34..207H}
}

@ARTICLE{Zhu2007,
   author = {{Zhu}, Z. and {Hartmann}, L. and {Calvet}, N. and {Hernandez}, J. and 
  {Muzerolle}, J. and {Tannirkulam}, A.-K.},
    title = "{The Hot Inner Disk of FU Orionis}",
  journal = {\apj},
archivePrefix = "arXiv",
   eprint = {0707.3429},
     year = 2007,
    month = nov,
   volume = 669,
    pages = {483-492},
      doi = {10.1086/521345},
   adsurl = {http://adsabs.harvard.edu/abs/2007ApJ...669..483Z}
}

@ARTICLE{Minniti2010,
   author = {{Minniti}, D. and {Lucas}, P.~W. and {Emerson}, J.~P. and {Saito}, R.~K. and 
  {Hempel}, M. and {Pietrukowicz}, P. and {Ahumada}, A.~V. and 
  {Alonso}, M.~V. and {Alonso-Garcia}, J. and {Arias}, J.~I. and 
  {Bandyopadhyay}, R.~M. and {Barb{\'a}}, R.~H. and {Barbuy}, B. and 
  {Bedin}, L.~R. and {Bica}, E. and {Borissova}, J. and {Bronfman}, L. and 
  {Carraro}, G. and {Catelan}, M. and {Clari{\'a}}, J.~J. and 
  {Cross}, N. and {de Grijs}, R. and {D{\'e}k{\'a}ny}, I. and 
  {Drew}, J.~E. and {Fari{\~n}a}, C. and {Feinstein}, C. and {Fern{\'a}ndez Laj{\'u}s}, E. and 
  {Gamen}, R.~C. and {Geisler}, D. and {Gieren}, W. and {Goldman}, B. and 
  {Gonzalez}, O.~A. and {Gunthardt}, G. and {Gurovich}, S. and 
  {Hambly}, N.~C. and {Irwin}, M.~J. and {Ivanov}, V.~D. and {Jord{\'a}n}, A. and 
  {Kerins}, E. and {Kinemuchi}, K. and {Kurtev}, R. and {L{\'o}pez-Corredoira}, M. and 
  {Maccarone}, T. and {Masetti}, N. and {Merlo}, D. and {Messineo}, M. and 
  {Mirabel}, I.~F. and {Monaco}, L. and {Morelli}, L. and {Padilla}, N. and 
  {Palma}, T. and {Parisi}, M.~C. and {Pignata}, G. and {Rejkuba}, M. and 
  {Roman-Lopes}, A. and {Sale}, S.~E. and {Schreiber}, M.~R. and 
  {Schr{\"o}der}, A.~C. and {Smith}, M. and {}, Jr., L.~S. and 
  {Soto}, M. and {Tamura}, M. and {Tappert}, C. and {Thompson}, M.~A. and 
  {Toledo}, I. and {Zoccali}, M. and {Pietrzynski}, G.},
    title = "{VISTA Variables in the Via Lactea (VVV): The public ESO near-IR variability survey of the Milky Way}",
  journal = {New Astronomy},
archivePrefix = "arXiv",
   eprint = {0912.1056},
     year = 2010,
    month = jul,
   volume = 15,
    pages = {433-443},
      doi = {10.1016/j.newast.2009.12.002},
   adsurl = {http://adsabs.harvard.edu/abs/2010NewA...15..433M}
}

@ARTICLE{Abraham2009,
       author = {{{\'A}brah{\'a}m}, P. and {Juh{\'a}sz}, A. and {Dullemond}, C.~P. and {K{\'o}sp{\'a}l}, {\'A}. and {van Boekel}, R. and {Bouwman}, J. and {Henning}, Th. and {Mo{\'o}r}, A. and {Mosoni}, L. and {Sicilia-Aguilar}, A. and {Sipos}, N.},
        title = "{Episodic formation of cometary material in the outburst of a young Sun-like star}",
      journal = {\nat},
         year = 2009,
        month = may,
       volume = {459},
       number = {7244},
        pages = {224-226},
          doi = {10.1038/nature08004},
archivePrefix = {arXiv},
       eprint = {0906.3161},
 primaryClass = {astro-ph.SR},
       adsurl = {https://ui.adsabs.harvard.edu/abs/2009Natur.459..224A}
}

@ARTICLE{Nayakshin2024,
       author = {{Nayakshin}, Sergei and {Elbakyan}, Vardan},
        title = "{On the origin of accretion bursts in FU Ori}",
      journal = {\mnras},
         year = 2024,
        month = feb,
       volume = {528},
       number = {2},
        pages = {2182-2198},
          doi = {10.1093/mnras/stae049},
archivePrefix = {arXiv},
       eprint = {2309.12072},
 primaryClass = {astro-ph.EP},
       adsurl = {https://ui.adsabs.harvard.edu/abs/2024MNRAS.528.2182N}
}

@ARTICLE{Nayakshin2024b,
       author = {{Nayakshin}, Sergei and {Cruz S{\'a}enz de Miera}, Fernando and {K{\'o}sp{\'a}l}, {\'A}gnes and {{\'C}alovi{\'c}}, Aleksandra and {Eisl{\"o}ffel}, Jochen and {Lin}, Douglas N.~C.},
        title = "{Episodic eruptions of young accreting stars: the key role of disc thermal instability due to Hydrogen ionization}",
      journal = {\mnras},
         year = 2024,
        month = may,
       volume = {530},
       number = {2},
        pages = {1749-1765},
          doi = {10.1093/mnras/stae877},
archivePrefix = {arXiv},
       eprint = {2403.04439},
 primaryClass = {astro-ph.SR},
       adsurl = {https://ui.adsabs.harvard.edu/abs/2024MNRAS.530.1749N}
}

@ARTICLE{Gaiadr3,
       author = {{Gaia Collaboration} and {Vallenari}, A. and {Brown}, A.~G.~A. and {Prusti}, T. and {de Bruijne}, J.~H.~J. and {Arenou}, F. and {Babusiaux}, C. and {Biermann}, M. and {Creevey}, O.~L. and {Ducourant}, C. and {Evans}, D.~W. and {Eyer}, L. and {Guerra}, R. and {Hutton}, A. and {Jordi}, C. and {Klioner}, S.~A. and {Lammers}, U.~L. and {Lindegren}, L. and {Luri}, X. and {Mignard}, F. and {Panem}, C. and {Pourbaix}, D. and {Randich}, S. and {Sartoretti}, P. and {Soubiran}, C. and {Tanga}, P. and {Walton}, N.~A. and {Bailer-Jones}, C.~A.~L. and {Bastian}, U. and {Drimmel}, R. and {Jansen}, F. and {Katz}, D. and {Lattanzi}, M.~G. and {van Leeuwen}, F. and {Bakker}, J. and {Cacciari}, C. and {Casta{\~n}eda}, J. and {De Angeli}, F. and {Fabricius}, C. and {Fouesneau}, M. and {Fr{\'e}mat}, Y. and {Galluccio}, L. and {Guerrier}, A. and {Heiter}, U. and {Masana}, E. and {Messineo}, R. and {Mowlavi}, N. and {Nicolas}, C. and {Nienartowicz}, K. and {Pailler}, F. and {Panuzzo}, P. and {Riclet}, F. and {Roux}, W. and {Seabroke}, G.~M. and {Sordo}, R. and {Th{\'e}venin}, F. and {Gracia-Abril}, G. and {Portell}, J. and {Teyssier}, D. and {Altmann}, M. and {Andrae}, R. and {Audard}, M. and {Bellas-Velidis}, I. and {Benson}, K. and {Berthier}, J. and {Blomme}, R. and {Burgess}, P.~W. and {Busonero}, D. and {Busso}, G. and {C{\'a}novas}, H. and {Carry}, B. and {Cellino}, A. and {Cheek}, N. and {Clementini}, G. and {Damerdji}, Y. and {Davidson}, M. and {de Teodoro}, P. and {Nu{\~n}ez Campos}, M. and {Delchambre}, L. and {Dell'Oro}, A. and {Esquej}, P. and {Fern{\'a}ndez-Hern{\'a}ndez}, J. and {Fraile}, E. and {Garabato}, D. and {Garc{\'\i}a-Lario}, P. and {Gosset}, E. and {Haigron}, R. and {Halbwachs}, J. -L. and {Hambly}, N.~C. and {Harrison}, D.~L. and {Hern{\'a}ndez}, J. and {Hestroffer}, D. and {Hodgkin}, S.~T. and {Holl}, B. and {Jan{\ss}en}, K. and {Jevardat de Fombelle}, G. and {Jordan}, S. and {Krone-Martins}, A. and {Lanzafame}, A.~C. and {L{\"o}ffler}, W. and {Marchal}, O. and {Marrese}, P.~M. and {Moitinho}, A. and {Muinonen}, K. and {Osborne}, P. and {Pancino}, E. and {Pauwels}, T. and {Recio-Blanco}, A. and {Reyl{\'e}}, C. and {Riello}, M. and {Rimoldini}, L. and {Roegiers}, T. and {Rybizki}, J. and {Sarro}, L.~M. and {Siopis}, C. and {Smith}, M. and {Sozzetti}, A. and {Utrilla}, E. and {van Leeuwen}, M. and {Abbas}, U. and {{\'A}brah{\'a}m}, P. and {Abreu Aramburu}, A. and {Aerts}, C. and {Aguado}, J.~J. and {Ajaj}, M. and {Aldea-Montero}, F. and {Altavilla}, G. and {{\'A}lvarez}, M.~A. and {Alves}, J. and {Anders}, F. and {Anderson}, R.~I. and {Anglada Varela}, E. and {Antoja}, T. and {Baines}, D. and {Baker}, S.~G. and {Balaguer-N{\'u}{\~n}ez}, L. and {Balbinot}, E. and {Balog}, Z. and {Barache}, C. and {Barbato}, D. and {Barros}, M. and {Barstow}, M.~A. and {Bartolom{\'e}}, S. and {Bassilana}, J. -L. and {Bauchet}, N. and {Becciani}, U. and {Bellazzini}, M. and {Berihuete}, A. and {Bernet}, M. and {Bertone}, S. and {Bianchi}, L. and {Binnenfeld}, A. and {Blanco-Cuaresma}, S. and {Blazere}, A. and {Boch}, T. and {Bombrun}, A. and {Bossini}, D. and {Bouquillon}, S. and {Bragaglia}, A. and {Bramante}, L. and {Breedt}, E. and {Bressan}, A. and {Brouillet}, N. and {Brugaletta}, E. and {Bucciarelli}, B. and {Burlacu}, A. and {Butkevich}, A.~G. and {Buzzi}, R. and {Caffau}, E. and {Cancelliere}, R. and {Cantat-Gaudin}, T. and {Carballo}, R. and {Carlucci}, T. and {Carnerero}, M.~I. and {Carrasco}, J.~M. and {Casamiquela}, L. and {Castellani}, M. and {Castro-Ginard}, A. and {Chaoul}, L. and {Charlot}, P. and {Chemin}, L. and {Chiaramida}, V. and {Chiavassa}, A. and {Chornay}, N. and {Comoretto}, G. and {Contursi}, G. and {Cooper}, W.~J. and {Cornez}, T. and {Cowell}, S. and {Crifo}, F. and {Cropper}, M. and {Crosta}, M. and {Crowley}, C. and {Dafonte}, C. and {Dapergolas}, A. and {David}, M. and {David}, P. and {de Laverny}, P. and {De Luise}, F. and {De March}, R. and {De Ridder}, J. and {de Souza}, R. and {de Torres}, A. and {del Peloso}, E.~F. and {del Pozo}, E. and {Delbo}, M. and {Delgado}, A. and {Delisle}, J. -B. and {Demouchy}, C. and {Dharmawardena}, T.~E. and {Di Matteo}, P. and {Diakite}, S. and {Diener}, C. and {Distefano}, E. and {Dolding}, C. and {Edvardsson}, B. and {Enke}, H. and {Fabre}, C. and {Fabrizio}, M. and {Faigler}, S. and {Fedorets}, G. and {Fernique}, P. and {Fienga}, A. and {Figueras}, F. and {Fournier}, Y. and {Fouron}, C. and {Fragkoudi}, F. and {Gai}, M. and {Garcia-Gutierrez}, A. and {Garcia-Reinaldos}, M. and {Garc{\'\i}a-Torres}, M. and {Garofalo}, A. and {Gavel}, A. and {Gavras}, P. and {Gerlach}, E. and {Geyer}, R. and {Giacobbe}, P. and {Gilmore}, G. and {Girona}, S. and {Giuffrida}, G. and {Gomel}, R. and {Gomez}, A. and {Gonz{\'a}lez-N{\'u}{\~n}ez}, J. and {Gonz{\'a}lez-Santamar{\'\i}a}, I. and {Gonz{\'a}lez-Vidal}, J.~J. and {Granvik}, M. and {Guillout}, P. and {Guiraud}, J. and {Guti{\'e}rrez-S{\'a}nchez}, R. and {Guy}, L.~P. and {Hatzidimitriou}, D. and {Hauser}, M. and {Haywood}, M. and {Helmer}, A. and {Helmi}, A. and {Sarmiento}, M.~H. and {Hidalgo}, S.~L. and {Hilger}, T. and {H{\l}adczuk}, N. and {Hobbs}, D. and {Holland}, G. and {Huckle}, H.~E. and {Jardine}, K. and {Jasniewicz}, G. and {Jean-Antoine Piccolo}, A. and {Jim{\'e}nez-Arranz}, {\'O}. and {Jorissen}, A. and {Juaristi Campillo}, J. and {Julbe}, F. and {Karbevska}, L. and {Kervella}, P. and {Khanna}, S. and {Kontizas}, M. and {Kordopatis}, G. and {Korn}, A.~J. and {K{\'o}sp{\'a}l}, {\'A}. and {Kostrzewa-Rutkowska}, Z. and {Kruszy{\'n}ska}, K. and {Kun}, M. and {Laizeau}, P. and {Lambert}, S. and {Lanza}, A.~F. and {Lasne}, Y. and {Le Campion}, J. -F. and {Lebreton}, Y. and {Lebzelter}, T. and {Leccia}, S. and {Leclerc}, N. and {Lecoeur-Taibi}, I. and {Liao}, S. and {Licata}, E.~L. and {Lindstr{\o}m}, H.~E.~P. and {Lister}, T.~A. and {Livanou}, E. and {Lobel}, A. and {Lorca}, A. and {Loup}, C. and {Madrero Pardo}, P. and {Magdaleno Romeo}, A. and {Managau}, S. and {Mann}, R.~G. and {Manteiga}, M. and {Marchant}, J.~M. and {Marconi}, M. and {Marcos}, J. and {Marcos Santos}, M.~M.~S. and {Mar{\'\i}n Pina}, D. and {Marinoni}, S. and {Marocco}, F. and {Marshall}, D.~J. and {Martin Polo}, L. and {Mart{\'\i}n-Fleitas}, J.~M. and {Marton}, G. and {Mary}, N. and {Masip}, A. and {Massari}, D. and {Mastrobuono-Battisti}, A. and {Mazeh}, T. and {McMillan}, P.~J. and {Messina}, S. and {Michalik}, D. and {Millar}, N.~R. and {Mints}, A. and {Molina}, D. and {Molinaro}, R. and {Moln{\'a}r}, L. and {Monari}, G. and {Mongui{\'o}}, M. and {Montegriffo}, P. and {Montero}, A. and {Mor}, R. and {Mora}, A. and {Morbidelli}, R. and {Morel}, T. and {Morris}, D. and {Muraveva}, T. and {Murphy}, C.~P. and {Musella}, I. and {Nagy}, Z. and {Noval}, L. and {Oca{\~n}a}, F. and {Ogden}, A. and {Ordenovic}, C. and {Osinde}, J.~O. and {Pagani}, C. and {Pagano}, I. and {Palaversa}, L. and {Palicio}, P.~A. and {Pallas-Quintela}, L. and {Panahi}, A. and {Payne-Wardenaar}, S. and {Pe{\~n}alosa Esteller}, X. and {Penttil{\"a}}, A. and {Pichon}, B. and {Piersimoni}, A.~M. and {Pineau}, F. -X. and {Plachy}, E. and {Plum}, G. and {Poggio}, E. and {Pr{\v{s}}a}, A. and {Pulone}, L. and {Racero}, E. and {Ragaini}, S. and {Rainer}, M. and {Raiteri}, C.~M. and {Rambaux}, N. and {Ramos}, P. and {Ramos-Lerate}, M. and {Re Fiorentin}, P. and {Regibo}, S. and {Richards}, P.~J. and {Rios Diaz}, C. and {Ripepi}, V. and {Riva}, A. and {Rix}, H. -W. and {Rixon}, G. and {Robichon}, N. and {Robin}, A.~C. and {Robin}, C. and {Roelens}, M. and {Rogues}, H.~R.~O. and {Rohrbasser}, L. and {Romero-G{\'o}mez}, M. and {Rowell}, N. and {Royer}, F. and {Ruz Mieres}, D. and {Rybicki}, K.~A. and {Sadowski}, G. and {S{\'a}ez N{\'u}{\~n}ez}, A. and {Sagrist{\`a} Sell{\'e}s}, A. and {Sahlmann}, J. and {Salguero}, E. and {Samaras}, N. and {Sanchez Gimenez}, V. and {Sanna}, N. and {Santove{\~n}a}, R. and {Sarasso}, M. and {Schultheis}, M. and {Sciacca}, E. and {Segol}, M. and {Segovia}, J.~C. and {S{\'e}gransan}, D. and {Semeux}, D. and {Shahaf}, S. and {Siddiqui}, H.~I. and {Siebert}, A. and {Siltala}, L. and {Silvelo}, A. and {Slezak}, E. and {Slezak}, I. and {Smart}, R.~L. and {Snaith}, O.~N. and {Solano}, E. and {Solitro}, F. and {Souami}, D. and {Souchay}, J. and {Spagna}, A. and {Spina}, L. and {Spoto}, F. and {Steele}, I.~A. and {Steidelm{\"u}ller}, H. and {Stephenson}, C.~A. and {S{\"u}veges}, M. and {Surdej}, J. and {Szabados}, L. and {Szegedi-Elek}, E. and {Taris}, F. and {Taylor}, M.~B. and {Teixeira}, R. and {Tolomei}, L. and {Tonello}, N. and {Torra}, F. and {Torra}, J. and {Torralba Elipe}, G. and {Trabucchi}, M. and {Tsounis}, A.~T. and {Turon}, C. and {Ulla}, A. and {Unger}, N. and {Vaillant}, M.~V. and {van Dillen}, E. and {van Reeven}, W. and {Vanel}, O. and {Vecchiato}, A. and {Viala}, Y. and {Vicente}, D. and {Voutsinas}, S. and {Weiler}, M. and {Wevers}, T. and {Wyrzykowski}, {\L}. and {Yoldas}, A. and {Yvard}, P. and {Zhao}, H. and {Zorec}, J. and {Zucker}, S. and {Zwitter}, T.},
        title = "{Gaia Data Release 3. Summary of the content and survey properties}",
      journal = {\aap},
         year = 2023,
        month = jun,
       volume = {674},
          eid = {A1},
        pages = {A1},
          doi = {10.1051/0004-6361/202243940},
archivePrefix = {arXiv},
       eprint = {2208.00211},
 primaryClass = {astro-ph.GA},
       adsurl = {https://ui.adsabs.harvard.edu/abs/2023A&A...674A...1G}
}

@article{Tennyson2016,
title = {The ExoMol database: Molecular line lists for exoplanet and other hot atmospheres},
journal = {Journal of Molecular Spectroscopy},
volume = {327},
pages = {73-94},
year = {2016},
note = {New Visions of Spectroscopic Databases, Volume II},
issn = {0022-2852},
doi = {https://doi.org/10.1016/j.jms.2016.05.002},
url = {https://www.sciencedirect.com/science/article/pii/S0022285216300807},
author = {Jonathan Tennyson and Sergei N. Yurchenko and Ahmed F. Al-Refaie and Emma J. Barton and Katy L. Chubb and Phillip A. Coles and S. Diamantopoulou and Maire N. Gorman and Christian Hill and Aden Z. Lam and Lorenzo Lodi and Laura K. McKemmish and Yueqi Na and Alec Owens and Oleg L. Polyansky and Tom Rivlin and Clara Sousa-Silva and Daniel S. Underwood and Andrey Yachmenev and Emil Zak},
}

@ARTICLE{Baraffe2015,
   author = {{Baraffe}, I. and {Homeier}, D. and {Allard}, F. and {Chabrier}, G.
  },
    title = "{New evolutionary models for pre-main sequence and main sequence low-mass stars down to the hydrogen-burning limit}",
  journal = {\aap},
archivePrefix = "arXiv",
   eprint = {1503.04107},
 primaryClass = "astro-ph.SR",
     year = 2015,
    month = may,
   volume = 577,
      eid = {A42},
    pages = {A42},
      doi = {10.1051/0004-6361/201425481},
   adsurl = {http://adsabs.harvard.edu/abs/2015A%26A...577A..42B}
}

@ARTICLE{Carvalho2024,
       author = {{Carvalho}, Adolfo S. and {Hillenbrand}, Lynne A.},
        title = "{Bolometric Corrections for FU Ori Object Accretion Disk Luminosities}",
      journal = {\apjl},
         year = 2024,
        month = nov,
       volume = {976},
       number = {1},
          eid = {L5},
        pages = {L5},
          doi = {10.3847/2041-8213/ad8cdf},
archivePrefix = {arXiv},
       eprint = {2410.22270},
 primaryClass = {astro-ph.SR},
       adsurl = {https://ui.adsabs.harvard.edu/abs/2024ApJ...976L...5C}
}

@ARTICLE{Szegedi-Elek2020,
       author = {{Szegedi-Elek}, E. and {{\'A}brah{\'a}m}, P. and {Wyrzykowski}, {\L}. and {Kun}, M. and {K{\'o}sp{\'a}l}, {\'A}. and {Chen}, L. and {Marton}, G. and {Mo{\'o}r}, A. and {Kiss}, C. and {P{\'a}l}, A. and {Szabados}, L. and {Varga}, J. and {Varga-Vereb{\'e}lyi}, E. and {Andreas}, C. and {Bachelet}, E. and {Bischoff}, R. and {B{\'o}di}, A. and {Breedt}, E. and {Burgaz}, U. and {Butterley}, T. and {Carrasco}, J.~M. and {{\v{C}}epas}, V. and {Damljanovic}, G. and {Gezer}, I. and {Godunova}, V. and {Gromadzki}, M. and {Gurgul}, A. and {Hardy}, L. and {Hildebrandt}, F. and {Hoffmann}, S. and {Hundertmark}, M. and {Ihanec}, N. and {Janulis}, R. and {Kalup}, Cs. and {Kaczmarek}, Z. and {K{\"o}nyves-T{\'o}th}, R. and {Krezinger}, M. and {Kruszy{\'n}ska}, K. and {Littlefair}, S. and {Maskoli{\={u}}nas}, M. and {M{\'e}sz{\'a}ros}, L. and {Miko{\l}ajczyk}, P. and {Mugrauer}, M. and {Netzel}, H. and {Ordasi}, A. and {Pak{\v{s}}tien{\.{e}}}, E. and {Rybicki}, K.~A. and {S{\'a}rneczky}, K. and {Seli}, B. and {Simon}, A. and {{\v{S}}i{\v{s}}kauskait{\.{e}}}, K. and {S{\'o}dor}, {\'A}. and {Sokolovsky}, K.~V. and {Stenglein}, W. and {Street}, R. and {Szak{\'a}ts}, R. and {Tomasella}, L. and {Tsapras}, Y. and {Vida}, K. and {Zdanavi{\v{c}}ius}, J. and {Zieli{\'n}ski}, M. and {Zieli{\'n}ski}, P. and {Zi{\'o}{\l}kowska}, O.},
        title = "{Gaia 18dvy: A New FUor in the Cygnus OB3 Association}",
      journal = {\apj},
         year = 2020,
        month = aug,
       volume = {899},
       number = {2},
          eid = {130},
        pages = {130},
          doi = {10.3847/1538-4357/aba129},
archivePrefix = {arXiv},
       eprint = {2005.11537},
 primaryClass = {astro-ph.SR},
       adsurl = {https://ui.adsabs.harvard.edu/abs/2020ApJ...899..130S}
}

@ARTICLE{Jorgensen2020,
       author = {{J{\o}rgensen}, Jes K. and {Belloche}, Arnaud and {Garrod}, Robin T.},
        title = "{Astrochemistry During the Formation of Stars}",
      journal = {\araa},
         year = 2020,
        month = aug,
       volume = {58},
        pages = {727-778},
          doi = {10.1146/annurev-astro-032620-021927},
archivePrefix = {arXiv},
       eprint = {2006.07071},
 primaryClass = {astro-ph.SR},
       adsurl = {https://ui.adsabs.harvard.edu/abs/2020ARA&A..58..727J}
}

@ARTICLE{Lucas2024,
       author = {{Lucas}, P.~W. and {Smith}, L.~C. and {Guo}, Z. and {Contreras Pe{\~n}a}, C. and {Minniti}, D. and {Miller}, N. and {Alonso-Garc{\'\i}a}, J. and {Catelan}, M. and {Borissova}, J. and {Saito}, R.~K. and {Kurtev}, R. and {Navarro}, M.~G. and {Morris}, C. and {Muthu}, H. and {Froebrich}, D. and {Ivanov}, V.~D. and {Bayo}, A. and {Caratti o Garatti}, A. and {Sanders}, J.~L.},
        title = "{The most variable VVV sources: eruptive protostars, dipping giants in the nuclear disc and others}",
      journal = {\mnras},
         year = 2024,
        month = feb,
       volume = {528},
       number = {2},
        pages = {1789-1822},
          doi = {10.1093/mnras/stad3929},
archivePrefix = {arXiv},
       eprint = {2401.14471},
 primaryClass = {astro-ph.SR},
       adsurl = {https://ui.adsabs.harvard.edu/abs/2024MNRAS.528.1789L}
}

@ARTICLE{Guo2024a,
       author = {{Guo}, Zhen and {Lucas}, P.~W. and {Kurtev}, R. and {Borissova}, J. and {Contreras Pe{\~n}a}, C. and {Yurchenko}, S.~N. and {Smith}, L.~C. and {Minniti}, D. and {Saito}, R.~K. and {Bayo}, A. and {Catelan}, M. and {Alonso-Garc{\'\i}a}, J. and {Caratti o Garatti}, A. and {Morris}, C. and {Froebrich}, D. and {Tennyson}, J. and {Mauc{\'o}}, K. and {Aguayo}, A. and {Miller}, N. and {Muthu}, H.~D.~S.},
        title = "{Spectroscopic confirmation of high-amplitude eruptive YSOs and dipping giants from the VVV survey}",
      journal = {\mnras},
         year = 2024,
        month = feb,
       volume = {528},
       number = {2},
        pages = {1769-1788},
          doi = {10.1093/mnras/stad3700},
archivePrefix = {arXiv},
       eprint = {2401.14464},
 primaryClass = {astro-ph.SR},
       adsurl = {https://ui.adsabs.harvard.edu/abs/2024MNRAS.528.1769G}
}

%

\begin{appendix}
\nolinenumbers
\section{SPHEREx data}
\label{sec:spherex}
The \textit{SPHEREx} spectrophotometry has a spatial resolution of $6".15$, covering a wavelength range from 0.75 to 5 $\mu$m. Two spectral epochs of SPICY~30021 were retrieved with observation dates around MJD 60891 and MJD 61094, during the outburst. Since no obvious variability ($>10\%$) is found between the two epochs, we combined the spectra to achieve full spectral coverage. Due to the low spatial resolution, SPICY~30021 is blended with the nearby companions, and one must remove the contribution from those companions to reveal the true flux of SPICY~30021. We inspected the spectral images and decided to use only the \textit{SPHEREx} data redder than 1 $\mu$m, as we have higher-resolution images from IMACS. In addition, the brightness of SPICY~30021 is less than 10\% of the companion sources at the short wavelength, which makes the de-blending unreliable.

%



According to the \textit{Gaia}~DR3 catalogue, the 3\arcsec~companion source is located at a distance of 2.6~kpc \citep{Gaiadr3}.  The BP-RP colour yields a $T_{\rm eff} \sim 8200$~K and an estimated foreground extinction of $A_V = 3.9$~mag. The extinction value is consistent with the $E(B-V) = 1.23$ provided by the Galactic Stellar Extinction Database with the given distance and coordinates \citep{Chen2019, Zhang2026}. We obtained the SED using photometric data spanning $0.44\text{--}5.63\,\mu\text{m}$, integrating optical measurements from \textit{Gaia} DR3 (including synthetic photometry) and VPHAS+ \citep{Drew2014}, NIR data from VISTA VVV, and mid-IR flux from \textit{Spitzer}. 

\begin{figure}[!h]
\centering
\includegraphics[width=\linewidth]{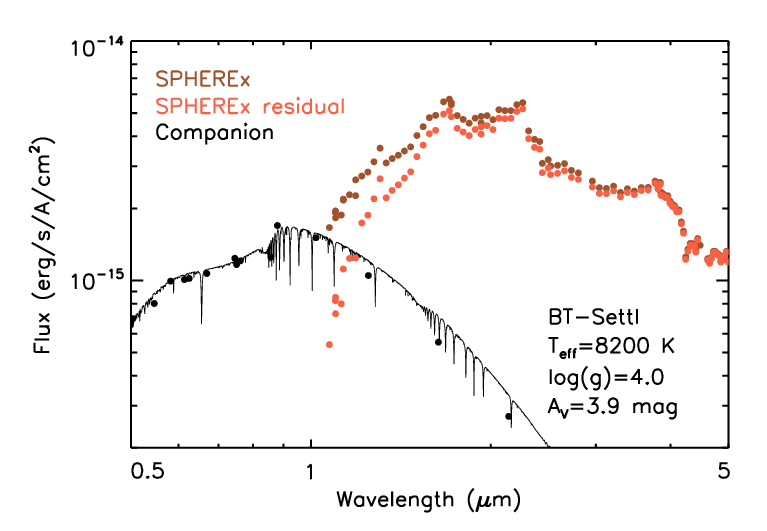}
\caption{The optical to near-IR SED of the 3\arcsec ~companion of SPICY30021 (black dots). We also present the blended \textit{SPHEREx} data (brown) and the spectrophotometry of SPICY30021 after removing the companion's flux (orange). In addition, we present a BT-Settl model, with $T_{\rm eff} = 8200$~K, $\log g = 4.0$ and extinction $A_V = 3.9$~mag.}
\label{fig:spherexsub}
\end{figure}

Adopting $A_V = 3.9$~mag and the stellar parameters from \textit{Gaia} DR3 ($T_{\rm eff}=8200$~K and $\log{g} = 4.0$), we generated a spectral model from BT-Settl \citep{Allard2001}. The model was then interpolated and subtracted from the \textit{SPHEREx} spectrum (see Fig.~\ref{fig:spherexsub}). To derive the integrated $K_{\rm s}$-band photometry, we applied the filter transmission curves and calculated the weighted mean of the $N$ valid spectral points within the corresponding bandpass ($F_0 = 4.27 \times 10^{-11}\,\mathrm{erg\,s^{-1}\,cm^{-2}\,\AA^{-1}}$). We then converted the weighted mean flux to the Vega magnitude using the standard zero-point obtained from the Spanish Virtual Observatory \citep{Rodrigo2020}\footnote{\url{https://svo2.cab.inta-csic.es/theory/fps/}}. Applying this methodology to the companion-subtracted residuals, we derived $K_s = 9.96 \pm 0.01$~mag, which is consistent with the NEWFIRM photometry $KX_s = 9.95 \pm 0.01$~mag.

\section{Distance measurements from proper motion and radial velocity}
\label{sec:distance}

SPICY~30021 is an optically obscured object, especially during its quiescent stage. Therefore, the \textit{Gaia} astrometry measurements are not available. Here, we applied the measured proper motions (PMs) from the VVV VIRAC2 catalogue to estimate the distance and to break the ambiguity between the near and far kinematic distance solutions  ($\rm PM_{RA} = -1.28 \pm 0.31 \,\rm mas\,yr^{-1}$ and $\rm PM_{Dec} = -0.83 \pm 0.36 \,\rm mas\,yr^{-1}$.). We used the same Monte-Carlo method described in \citet{Guo2021}, using the Milky Way disk rotation and peculiar velocity dispersion models from \citep{Bensby2013}. For an input set of PMs, a distance solution is computed by randomly sampling 1 million stars following the peculiar velocity distribution. We also considered a Gaussian dispersion of PMs based on the errors from PM measurements. A solution for SPICY~30021 is presented in Fig.~\ref{fig:distance_pm}. The highest possibility is located around 1.6~kpc. 

\begin{figure}[!h]
\centering
\includegraphics[width=\linewidth]{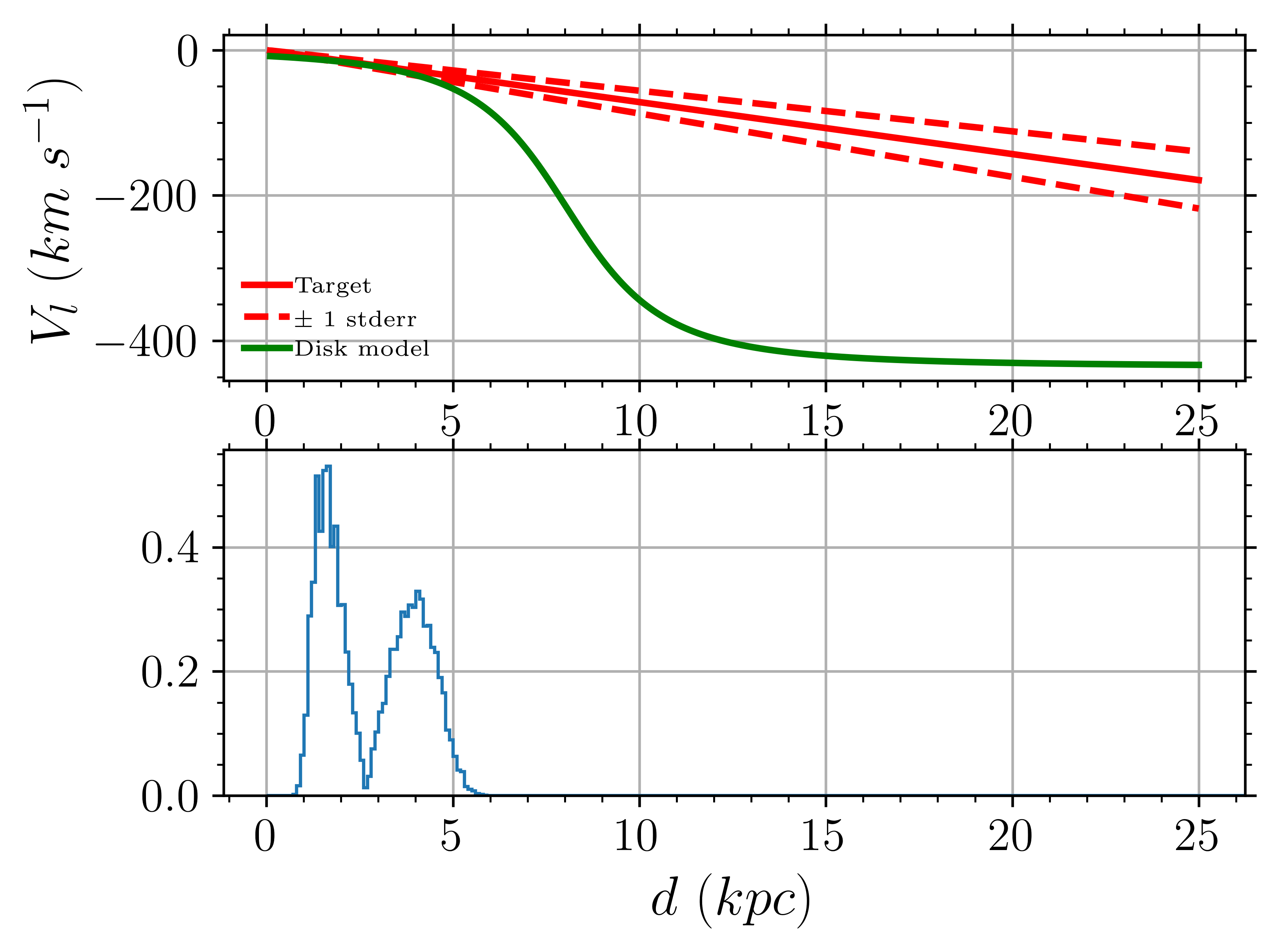}
\caption{\textit{Upper:} Galactic tangential velocity at $gl = 327.8^\circ$, as a function of distance. \textit{Bottom:} Distance solution with given PMs and gl.}
\label{fig:distance_pm}
\end{figure}

We also derived the kinematic distance of SPICY ~30021 using the radial velocity measured from the CO bandhead fitting ($\rm V_{LSR} = -26.5~km~s^{-1}$), and assume that CO is from the circumstellar gas disk instead of from a blueshifted wind. Limited by the spectral resolution and pixel scale, the uncertainty of the RV measurement is on the order of 5~km~s$^{-1}$. We used the Monte-Carlo online tool\footnote{\url{https://www.treywenger.com/kd/index.php}} designed in \citet{Wenger2018}, which considers the uncertainties from the RV measurement and the Galactic rotation curve \citep{Reid2014}. The detailed parameters applied for the online tool were introduced in \citet{Guo2021}. The near kinematic distance solution is $d_k = 1.7\pm0.4$~kpc. 

Using the high-angular-resolution extinction map provided in \citet{Zucker2025}, we illustrated the line-of-sight extinction towards the Galactic locations of SPICY~30021. In Fig.~\ref{fig:extinction_distance}, the first peak of $E(B-V)$ appears between 1.6 and 1.8~kpc, indicating the location of the Sagittarius-Carina arm. The interstellar extinction at a distance of 1.7~kpc is $A_V = R_V \times E(B-V) = 2.5$~mag, {proving} that most of the extinction measured {in} SPICY~30021 is from the circumstellar material or from an unresolved small-scale cloud.

\begin{figure}
\centering
\includegraphics[width=\linewidth]{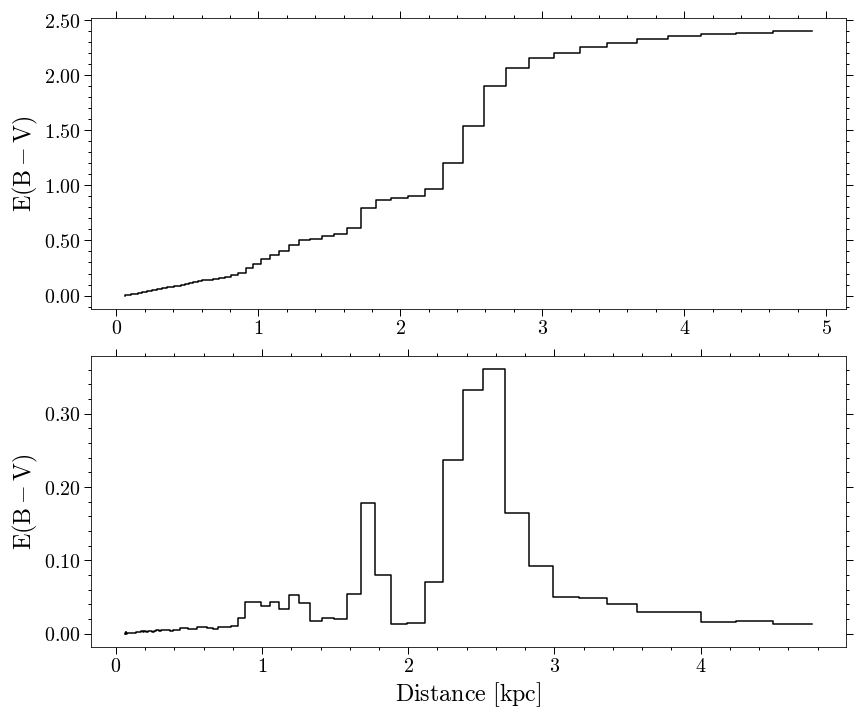}
\caption{Accumulated and local dust extinction as a function of distance. We applied the direction of SPICY~30021 to the extinction map from \citet{Zucker2025}.}
\label{fig:extinction_distance}
\end{figure}

\section{Near-infrared spectral features}
\label{sec:line}

In Fig.~\ref{fig:extinction}, we show the dereddened NIR spectra of SPICY~30021 obtained by TSpec with $A_V$ ranging between 10.5 and 14.5 mag, compared with FU Ori observed on IRTF. This plot illustrates the accuracy of our extinction measurement. The spectral slope between the $J$ and $H$ bands changes significantly with different $A_V$, and the best-matched value is 12.5 mag. After visual inspection, the error bar should be less than 0.5 mag.

In Fig.~\ref{fig:spec_features}, we present the NIR spectra from SOAR/TSpec and VLT/KMOS. Individual spectral features are presented on the plot as labels. Water vapour and CO molecular absorption bands are seen beyond 2.29~$\mu$m. For the water vapour absorption band, the slope presented in Fig.~\ref{fig:spec_features} is the slope we used to normalise the CO bandhead feature for the model fitting. 

\begin{figure}
    \centering
    \includegraphics[width=0.98\linewidth]{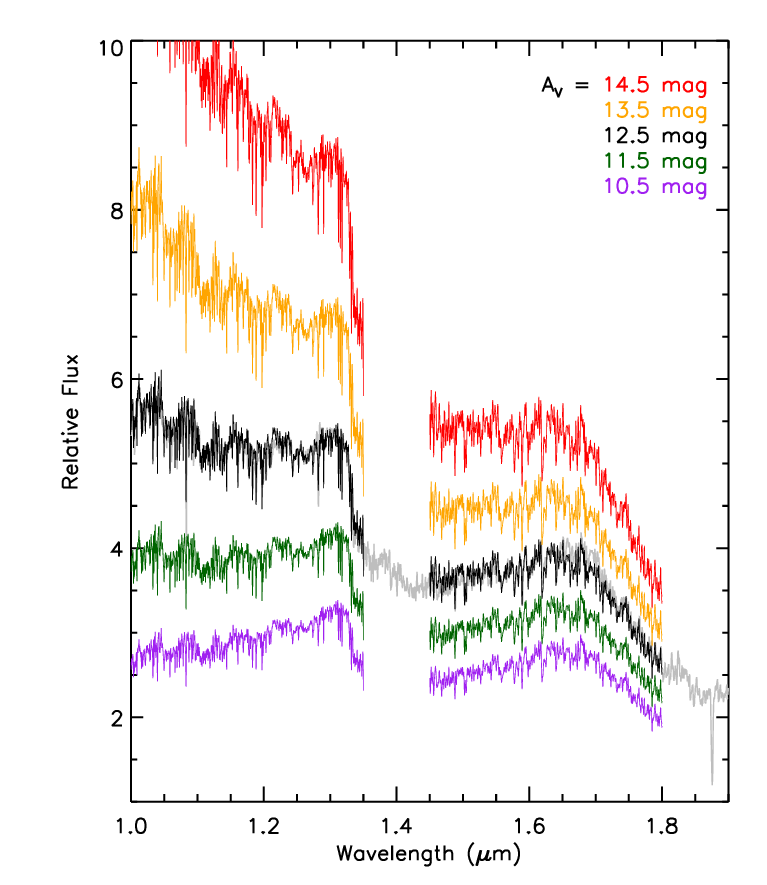}
    \caption{Dereddened spectra of SPICY~30021 with different $A_V$ ranging from 10.5 to 14.5 mag. The dereddened spectrum of FU Ori is presented in grey, matching the $A_V = 12.5$ mag spectrum of SPICY~30021.}
    \label{fig:extinction}
\end{figure}

\begin{figure*}
    \centering  \includegraphics[width=0.98\linewidth]{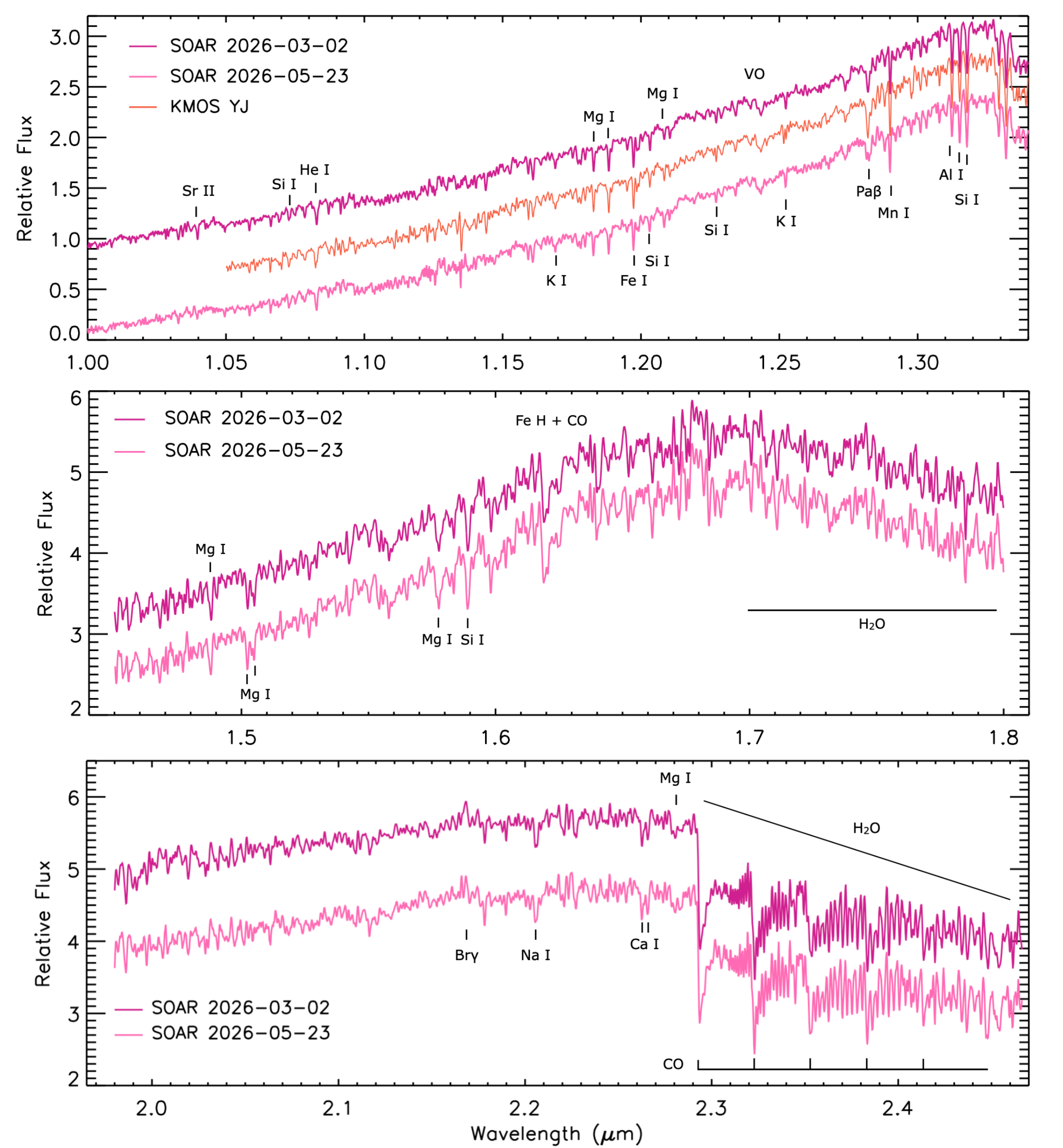}
    \caption{Near-IR spectra of SPICY~30021 obtained during the eruptive stage. Specific spectral features, including absorption lines and molecular bands, are labelled on the plot.}
    \label{fig:spec_features}
\end{figure*}
\end{appendix}

\end{document}